\documentclass[aps,prx,twocolumn,showpacs,floatfix,superscriptaddress]{revtex4-1}

\usepackage{amssymb,amsmath,amstext}                
\usepackage{graphicx}                                               
\usepackage{epstopdf}                                               
\usepackage{color}                                                     
\usepackage{bm}                                                        
\usepackage{appendix}                                              
\usepackage[utf8]{inputenc}
\usepackage{bbold}
\usepackage{bbm}
\usepackage{ulem}
\usepackage{latexsym}
\usepackage{xcolor}
\usepackage{braket}
\usepackage{verbatim}
\definecolor{lblue} {RGB}{51,71,158}
\usepackage[colorlinks=true,citecolor=blue,linkcolor=blue,urlcolor=lblue]{hyperref}

\DeclareMathOperator{\Tr}{Tr}

\newcommand{\average}[1]{\overline{#1}}

\begin{document}

\title{Many-body localization in one dimensional optical lattice with speckle disorder
}
\author{Artur Maksymov}
\affiliation{Institute of Theoretical Physics, Jagiellonian University in Krakow,  \L{}ojasiewicza 11, 30-348 Krak\'ow, Poland }

\author{Piotr Sierant}
\affiliation{Institute of Theoretical Physics, Jagiellonian University in Krakow, \L{}ojasiewicza 11, 30-348 Krak\'ow, Poland }
\affiliation{ICFO- Institut de Sciences Fotoniques, The Barcelona Institute of Science and Technology, Av. Carl Friedrich Gauss 3, 08860 Castelldefels (Barcelona), Spain}

\author{Jakub Zakrzewski}
\affiliation{Institute of Theoretical Physics, Jagiellonian University in Krakow,   \L{}ojasiewicza 11, 30-348 Krak\'ow, Poland }
\affiliation{Mark Kac Complex
Systems Research Center, Jagiellonian University in Krakow, 30-348 Krak\'ow,
Poland. }
\email{jakub.zakrzewski@uj.edu.pl}

\date{\today}

                              
\begin{abstract}
 
 The many-body localization transition for Heisenberg spin chain with a speckle disorder is studied. Such a model is
equivalent to a system of spinless fermions 
in an optical lattice  with an additional speckle
field.
Our numerical results show that the many-body localization transition in speckle disorder falls within the
same universality class as the transition in an uncorrelated random disorder, in contrast to the
quasiperiodic potential typically studied in experiments.
This hints at possibilities of experimental studies of the role of rare Griffiths regions and of the 
interplay of ergodic and localized grains at the many-body localization transition.
Moreover, the speckle potential allows one to study the role of correlations in disorder
on the transition.
We study both spectral and 
dynamical properties of the system 
focusing on observables that are sensitive to the disorder type and its correlations.
In particular, distributions of local imbalance at long 
times provide an experimentally available tool that reveals 
the presence of small ergodic grains even deep in the many-body localized
phase in a correlated speckle disorder. 
\end{abstract}

\maketitle

\section{Introduction}

Isolated quantum many-body systems are generically expected to reach thermal equilibrium 
according to eigenstate thermalization hypothesis \cite{Deutsch91, Srednicki94,Rigol08, Alessio16}.
The approach to thermal equilibrium may be precluded by 
strong disorder resulting in phenomenon of Many-Body Localization (MBL) 
\cite{Gornyi05, Basko06}
investigated in recent years both theoretically and experimentally
(for reviews see \cite{Huse14, Nandkishore15, Alet18, Abanin19}).
Further examples of
nonergodic many body-systems {include} models with 
  constrains 
\cite{Sala20,Khemani20, Rakovszky20},  lattice gauge theories \cite{Smith17, Brenes18, Magnifico20, 
Chanda20, Giudici20, Surace20} 
often linked with periodic oscillatory behavior coined, 
in the wake of well known quantum chaos notion \cite{Heller84,Bogomolny88}, as quantum 
scarring \cite{Turner18, Wen19, Khemani19, Iadecola19, Iadecola19a}, as well as surprizingly basic systems
as interacting particles in tilted lattices (Stark-like localization)
\cite{Schulz19,vanNieuwenburg19} or even harmonic potentials featuring coexistence of localized 
and delocalized phases \cite{James19,Chanda20c,Yao20b}.

The theoretical studies of MBL typically consider uniform uncorrelated random potential 
as a source of disorder in the system. In contrast,
the experimental setups used much easier to realize quasiperiodic
potential \cite{Schreiber15,Choi16} correlated at arbitrary length scales.
The behavior of one-dimensional (1D) models deep in the 
localized phase is similar in both cases
(leading e.g. to preservation of the information about initial states in time dynamics 
\cite{Schreiber15,Choi16}). The situation is more complicated in the crossover between
localized and extended phases. It is claimed even that the observed behavior suggest 
different universality classes of MBL transition depending on the disorder {type} \cite{Khemani17,Sierant19b}.
For uncorrelated disorder one expects to observe  the influence of rare 
events, the so called Griffiths regions \cite{Griffiths69,Vojta10} i.e.
grains of the minority phase on the either side of the transition (e.g. 
presence of ergodic grains on the localized side).
Those affect the time dynamics and lead to e.g. subdiffusive transport on the delocalized side
\cite{Agarwal15,Luitz15,Agarwal17,Gopalakrishnan16,Luitz16,Luitz17b}. 
Even though the rare Griffiths regions are a priori absent in the deterministic quasiperiodic potential,
the resulting dynamics are similar as in the uncorrelated disorder featuring a power-law decay of
time-correlators as well as a power-law growth
of the entanglement entropy \cite{Luitz16, Luschen17} - for a recent review see 
\cite{Gopalakrishnan20}.

Surpizingly much less is known about MBL in {a} random speckle potential, despite the fact that
such a potential has been successfully used {in} single particle physics e.g. 
for the experimental demonstration of Anderson localization in cold atomic gases \cite{Billy08,Piraud13}.
For attractively interacting bosons the bright soliton can be trapped in a speckle-disorder
potential and get Anderson localized 
\cite{Sacha09,Delande13}. 
{A study of MBL in two-dimensional continuum \cite{Nandkishore14} concludes that perturbation theory diverges for arbitrarily 
weak interactions in a speckle potential.
Moreover, it is not clear whether the insulating state of a strongly correlated atomic Hubbard gas 
in a speckle potential observed in center-of-mass velocity measurements \cite{Kondov14} can be attributed
to MBL since the phenomenon is believed to be not stable beyond one dimension \cite{DeRoeck17}.
} {A} recent  theoretical  study {of MBL in a speckle potential}
 \cite{Mujal19} was limited to 
few particles only due to numerical requirements of the continuum approach.

\begin{figure}
\includegraphics[width=0.95\linewidth]{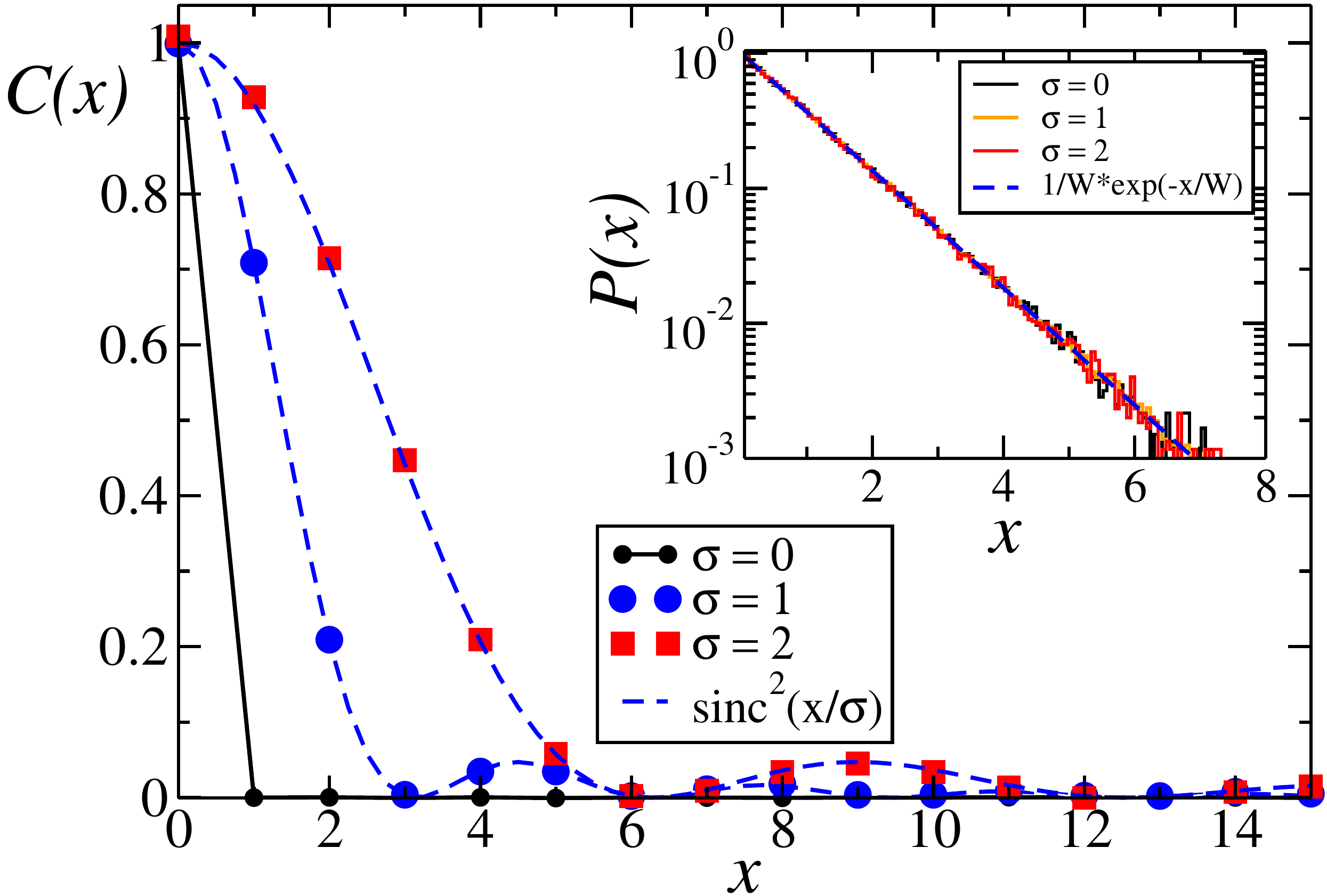}
\caption{The speckle distribution (inset) and its correlation function 
for disorder strength $W=1$ and different correlation lengths $\sigma$ as
indicated.
}
\label{fig:speckle}
\end{figure}

The aim of this work is to provide an in depth study of MBL in a speckle potential
in a one dimensional chain which is the typical geometry
in which MBL is studied both experimentally and theoretically. 
While MBL has
been studied for spinless or spin-1/2 fermions \cite{Mondaini15}
as well as for bosons in optical lattices 
potential \cite{Sierant17, Sierant18, Orell19, Hopjan19, Yao20} we choose to consider the simplest, 
paradigmatic model used in MBL studies,
namely, the disordered Heisenberg chain. There are at least two reasons underlying this choice.
Firstly, for random uniform as well as quasiperiodic disorder the Heisenberg spin chain has
been quite deeply analyzed already, thus a direct comparison with random uniform results 
allow us to better comprehend the differences resulting from the nature of the speckle potential.
Secondly, with the local on-site Hilbert space dimension equal to 2, one may, using exact
diagonalization approach reach system size of the order of $L=20$ straightforwardly. That allows us 
for an in-depth analysis of properties of the system. That would be much harder 
for bosons  or spinful fermions -- for the latter, in addition,  the 
inherent SU(2) symmetry affects deeply
the MBL transition \cite{Prelovsek16, Zakrzewski18, Kozarzewski18,Protopopov19,Suthar20}.

The paper is organized as follows. Section~\ref{model} introduces the model and the 
speckle disorder, we review its basic properties there. With this knowledge we consider 
 properties of {eigevalues and eigenstates of }
 the model in Section~\ref{spect} while the time dynamics is discussed
in Section~\ref{time}. Appendices provide additional discussion on specialized topics. 
We summarize our findings in Section~\ref{concl}.

\section{The model. }
\label{model}

We consider a 1D Heisenberg spin chain, widely used in MBL studies
\cite{Luitz15,Agarwal15, Bera15, 
Bera17, Herviou19, 
Colmenarez19}.
This model maps, via Jordan-Wigner transformation, to a 
system of interacting spinless fermions 
which allows us to make the connection with 
optical lattice experiments.
Instead of quasiperiodic disorder imposed 
by a secondary weak optical lattice with period incommensurate with the 
primary lattice (in which the tight binding approximation is inherently
assumed) as in experiments \cite{Schreiber15,Luschen17}  we imagine that
the  disorder is added by an additional optical potential due to a speckle radiation. 
This additional light may operate on a different optical transition than 
the primary optical lattice and, within the tight binding approximation
for the latter, leads to a desired speckle disorder. The resulting Hamiltonian
of the system reads:
\begin{equation}
 H= J\sum_{i=1}^{L-1} \ \vec{S}_i \cdot \vec{S}_{i+1} + \sum_{i=1}^{L} h_i S^z_i,
 \label{eq: XXZ}
\end{equation}
where  $\vec{S}_i$ are spin-1/2 matrices, $J=1$ will be considered as the energy unit,
open boundary conditions are assumed. {The local magnetic fields,} $h_i$,
are drawn from the speckle 
distribution $P(x)=\frac{1}{W}exp(-x/W);\ x >0$ where $\average x=W$ (an overbar 
denotes an average over disorder realizations). Similarly, $W$ is the standard 
deviation of the exponential distribution. Importantly, $h_i$ values may be  
correlated depending on their relative position. 
The speckle is typically generated by transmission of light 
through a ground glass plate \cite{Goodmanbook}. The correlations
in the speckle pattern result from interference of light scattered by 
different parts of the plate and are, therefore, controlled by the aperture of the
object. Assuming a rectangular plate, see \cite{Lugan09} for more details, 
the correlation function takes a form  (in discrete representation)
$\average{h_ih_j}=W^2C(|i-j|/\sigma)$ with $C(y)=[\sin(y)/y]^2$ and 
$|i-j|$ the distance (unit lattice constant assumed) - compare Fig.~\ref{fig:speckle}.

Few remarks are in order.  The disorder is asymmetric with assumed positive $x$. 
We could also change the sign of all $h_i$ to have an opposite case (in atomic 
implementation a change of the sign corresponds to the change of the sign of the
detuning on the atomic transition). This sign is relevant and important for low
lying states \cite{Sacha09,Delande13} (as the disorder corresponds to either
peaks or valleys of the potential). However we shall consider highly excited 
states from the middle of the spectrum and this sign becomes irrelevant. 
Second, in an optical implementation
$\sigma$ may be as low as $0.26\mu m$ \cite{Billy08} i.e. a fraction of the 
typical lattice spacing in experiment \cite{Schreiber15,Luschen17}. In the 
tight binding model we have then simply an uncorrelated disorder. Increasing
$\sigma$ we may study how finite correlations in the potential affect MBL, 
the option apparently not available for other types of experimentally relevant
disorder used till now.

{For reference, we shall use also the uniform random (UR) disorder, for which 
the fields $h_i$ are independent random variables drawn from uniform
distribution on interval $[-W,W]$. }

\section{Properties of eigenvalues and eigenstates}
\label{spect}

\subsection{Locating the transition}

With the model defined we study first its spectral properties 
to verify the presence of the ergodic-MBL transition.
Consider a mean 
gap ratio, 
$\average r$,  {calculated as an average of}
\begin{equation}
r_i = \min \{ \frac{s_{i+1}}{s_i}, \frac{s_i}{s_{i+1}} \},
\label{gap}
\end{equation}
where $s_i=E_{i+1}-E_i$ (with $E_i$ being eigenvalues of the system)
and the average is taken  over a region of spectrum and over disorder realizations. 
The mean gap ratio was proposed  as a simple 
probe of level statistics in \cite{Oganesyan07} with $\average r\approx 0.38$ for Poisson statistics (PS) 
(corresponding to localized, integrable case)
and $\average r\approx 0.53$ for 
 Gaussian orthogonal  ensemble (GOE) \cite{Mehtabook,Haakebook} of random matrices well describing statistically
an ergodic system.

\begin{figure}[b]
	\includegraphics[width=0.9\linewidth]{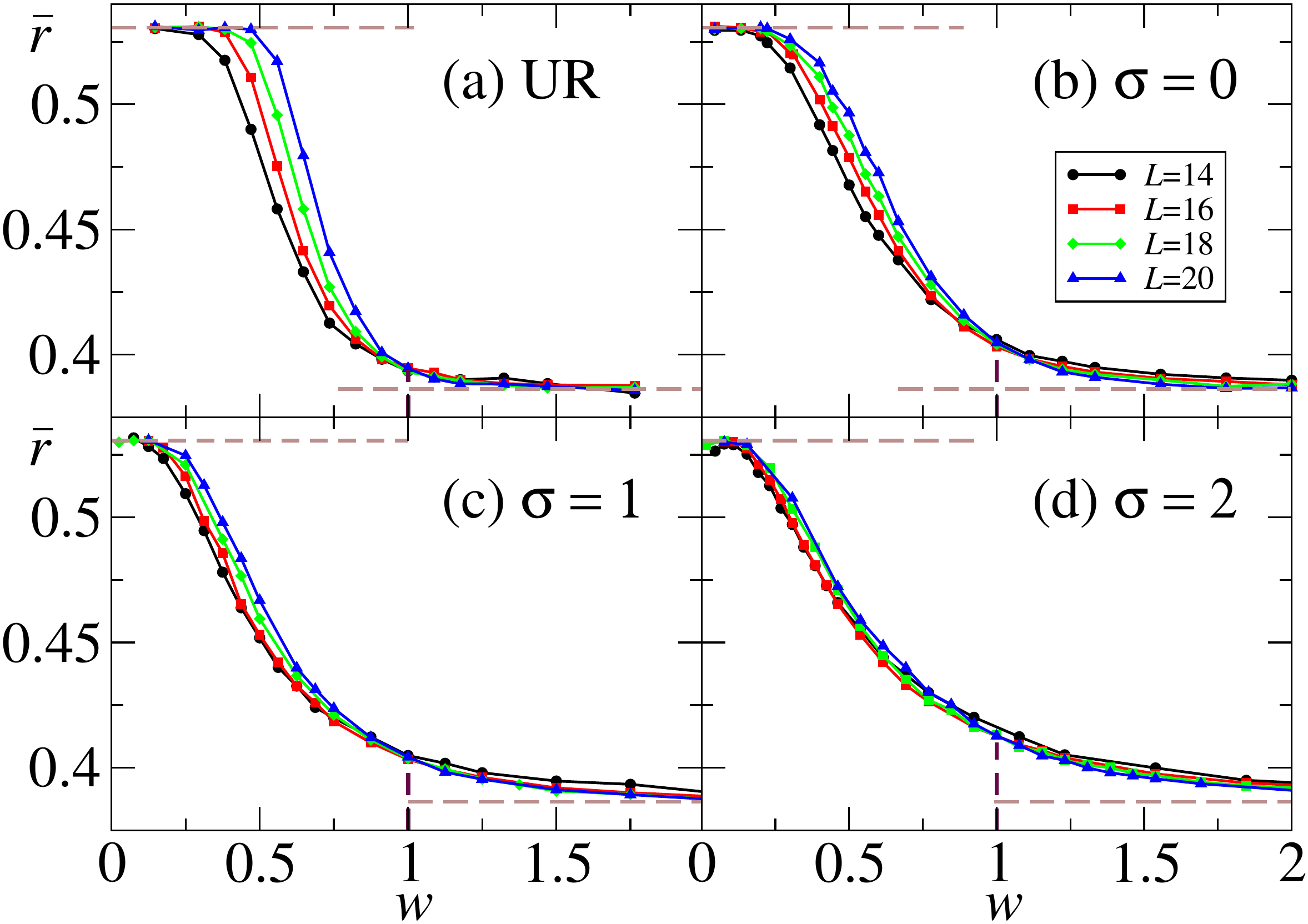}
	\caption{Mean gap ratio as a function of the rescaled disorder amplitude $w = W/W_c$ -- 
	for $W_c$ values please see Table~I. Panel (a) shows the reference behavior for a 
	random uniform disorder while the remaining panels correspond to different scale of
	correlations {of the speckle potential}
	as indicated by $\sigma$ values in the panels.  The transition in a 
	speckle potential seems much broader {with} smaller differences between curves for different 
	system sizes. Increasing correlations make this effect much stronger.}
	\label{rbar2}
\end{figure}

We determine the mean gap ratio as a function of the disorder amplitude $W$ for several system sizes. Since the total
spin projection on the $z$ axis is conserved, we consider the largest nontrivial sector of the Hamiltonian with $\sum S_i^z=0$ 
-- that restricts system sizes considered to $L$ even. Typically, we consider about $N=300$ 
eigenenergies from the middle of the spectrum, i.e. $\varepsilon\approx0.5$ where 
 $\varepsilon=(E-E_{min})/(E_{max}-E_{min})$
with $E_{min} (E_{max})$ being the lowest (highest) eigenvalue for a given disorder realization.
The results are averaged over 1000 
disorder realizations or twice that number close to the estimated transition point. 
Data for $L=14$ and $L=16$ are obtained by full exact diagonalization, the ones for $L=18$ and $L=20$ are obtained
using POLFED algorithm \cite{Sierant20p}. The results are shown in 
Fig.~\ref{rbar2}. Curves corresponding to different system sizes cross
typically in the vicinity of $\average r\approx 0.4$ - this crossing point is taken as an estimate 
of the critical disorder value, $W_c$ for different disorders. We refrain from using the procedure of a single
parameter finite size scaling  \cite{Luitz15,Mondaini15,Khemani17} as it has became apparent recently
\cite{Laflorencie20,Suntajs20} that the transition may be of Kosterlitz-Thouless type and 
such a finite size scaling approach may be not valid.
  
  The critical values of disorder for  $\varepsilon \approx 0.5$ for all considered models are given in Table~I. 
  Note that for {UR}  disorder we get 
  $W_c\approx3.4$ close to $W_c\approx 3.3$ obtained with finite size scaling {in the
  system with open boundary conditions} \cite{Chanda20t}. 
  The critical disorder values are used
  to rescale the disorder amplitude $w=W/W_c$ to facilitate a comparison between various
  models of disorder. From 
  now on we shall use, whenever possible, the rescaled disorder, $w$.  
\begin{table}[]
	\caption{The critical values of disorder for different disorder types considered in this work  obtained {at the middle of the spectrum}, $\varepsilon=0.5$.}
	\begin{tabular}{|c|c|c|c|c|c|}
		\hline
		& {UR}	& $\sigma=0$    & $\sigma=1$    & $\sigma=2$       \\ \hline
		$W_{c}$ & 3.4		& 2.25        & 4           & 6.5           \\ \hline
	\end{tabular}
\end{table}

Comparison of Fig.~\ref{rbar2}(a) and Fig.~\ref{rbar2}(b) 
shows that the crossover between ergodic and localized 
phases for UR disorder is sharper than for the
uncorrelated speckle disorder ($\sigma=0$). The disorder strength at which the average
gap ratio $\overline r$ departs from the GOE value reaches $w\approx 0.5$ for the UR
disorder and system size $L=20$ and $w\approx 0.25$  for speckle disorder with $\sigma=0$. 
This can be traced back to the unbounded on-site distribution of the speckle potential: the
probability of having a fully ergodic system at e.g. $w=0.3$ is diminished, in comparison to the 
UR disorder case, by the rare events in which the field $h_i$ is large {on one of the sites}. 
The effect of broadening of the crossover is further enhanced 
when the correlations ($\sigma>0$) are introduced in the speckle 
potential.
 
 This suggests that finite size 
effects are stronger for the correlated disorder
{which}
is seemingly contradicted 
by the fact that curves for different sizes $L$ are much closer to each other 
for the {correlated} speckle disorder.
{However, a} weaker size dependence {of $\bar r$} 
is a sign of strong finite size effects
-- relatively small changes in system sizes $L$ available to exact diagonalization are simply 
too small  to have a visible effect on $\average r$ dependence for the correlated speckle disorder.

\subsection{Inter-sample randomness}

\begin{figure}[b]
	\includegraphics[width=0.9\linewidth]{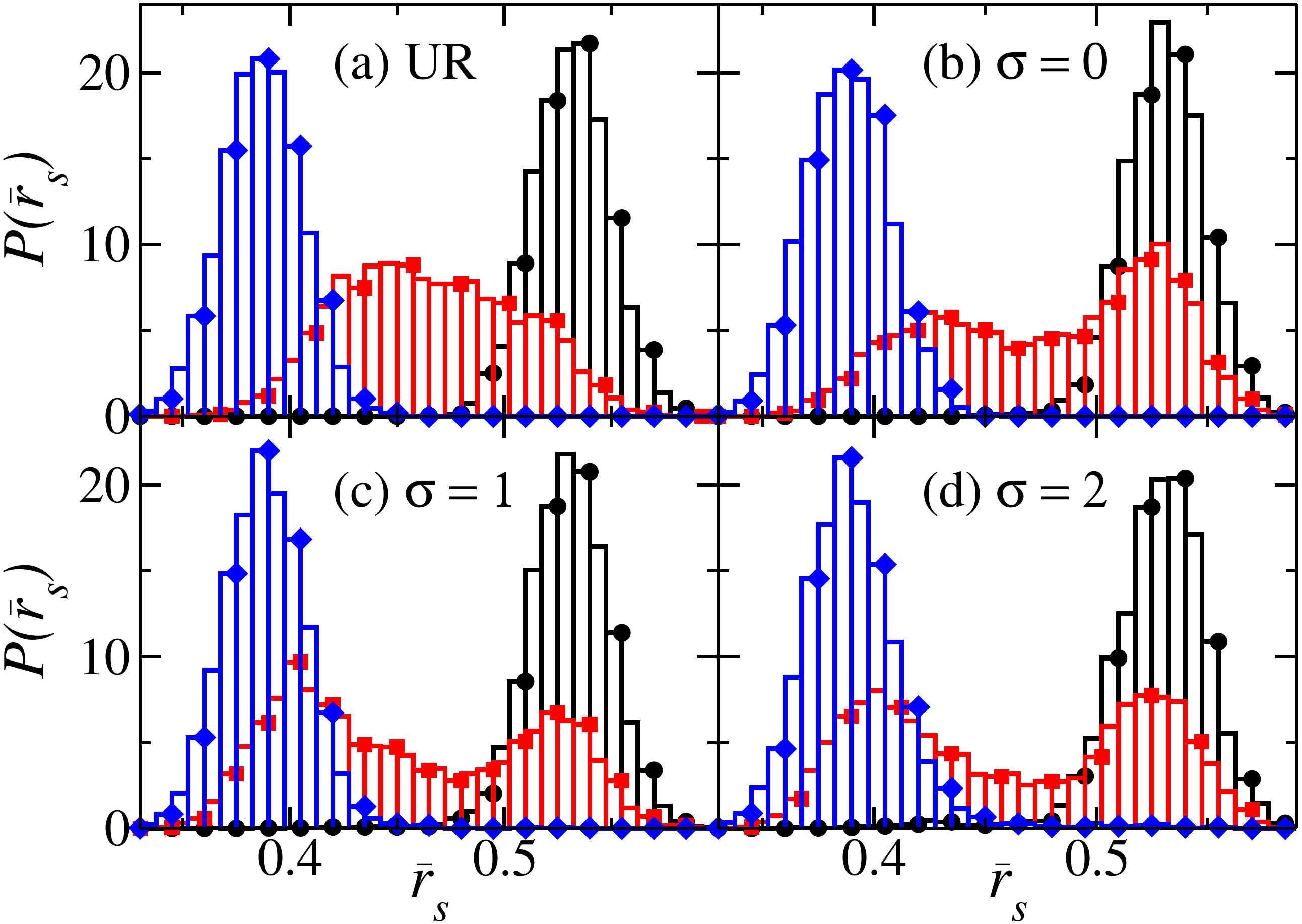}
	\caption{{Distributions {$P(\overline r_S)$} of 
	{the sample-averaged gap ratio} obtained  for single 
	realizations of disorder for random (a) and speckle potential with $\sigma=0$ 
	(b), $\sigma=1$ (c) and $\sigma=2$ (d). The black circles and blue diamonds are 
	for delocalized and localized phase with $w = 0.15$ and $w = 2.35$, respectively;
	the red circles are for transient regime with $w = 0.6$ for random potential and 
	$w = 0.5$ for {the} speckle. The data are obtained for the system size $L=16$, 4500 
	realizations and 300 levels around $\varepsilon=0.5$.}}
	\label{fig:rs}
\end{figure}

A more detailed characteristic of ergodic to MBL transition 
is obtained when one examines variation of system properties for individual disorder realizations 
\cite{Sierant19b}.
To that end we average the gap ratios $r_i$ obtained from $300$ eigenvalues 
from the middle of the spectrum ($\varepsilon=0.5$) for a single disorder
realization which yields a sample-averaged gap ratio $\bar{r}_s$. 
The distributions of $\bar{r}_s$ 
(obtained from calculating $\bar{r}_s$ for many disorder realizations)
vastly differ between UR disorder and quasiperiodic potential
\cite{Sierant19b} pointing out 
the dominant role of inter-sample randomness in the MBL transition in the UR disorder.
The {large inter-sample randomness}, in turn, may be {linked}
to the existence of rare Griffiths regions.
Hence, it was suggested \cite{Khemani17} that MBL
transitions for UR disorder and quasiperiodic 
potential belong to different universality classes.

The distribution of the sample averaged gap ratio
$P(\bar{r}_s)$ obtained for the speckle disorder
are compared with the UR disorder case  
in Fig.~\ref{fig:rs}. Three values of the rescaled
disorder are considered:
$w = 0.15$ and $w=2.35$ corresponding to
delocalized and localized phases, respectively, and intermediate one corresponding 
to $w = 0.5$ (for RU disorder  $w = 0.6$). Such intermediate values of 
disorder correspond to the maximal inter-sample randomness in the system with the biggest 
variance of the $\bar r_s$ distribution and in the thermodynamic limit they 
tend to the respective critical values $W_c$ 
provided that  MBL persists in 
the thermodynamic limit (this issue is a topic of the current debate
 \cite{Suntajs19,Panda20,Sierant20b,Sierant20p,Suntajs20}).

Observe that, as it could be expected, the distributions for {the} localized and 
 delocalized cases are quite similar for {UR}
and for the speckle
disorders with different correlation lengths. The situation is markedly different in the transition regime.
For the system with {UR disorder} the $P(\bar{r}_s)$ distribution is
unimodal, although broad.  On the other hand, for the speckle disorder one can observe a
bimodal shape. 
With the increase of the correlation length $\sigma$ the two peaks observed 
for speckle potential become more pronounced and overlap more the distributions 
for delocalized and localized phases.

Apparently, the sample averaged gap ratio distribution catches
an important difference
between the speckle and UR disorder. The speckle distribution is exponential favoring 
small values of disorder. Thus, it is quite probable to obtain nearby sites with
very small difference in the local potential -- that facilitates transport. 
This observation {may be}
quantified by finding the probability distribution for the difference
between consecutive random numbers which, for speckle disorder is also exponential.
At the intermediate rescaled disorder, $w=0.5$, the probability of having an ergodic 
system $(\bar{r}_s\approx0.53)$ is enhanced for the uncorrelated speckle disorder. At the
same time, the unbounded exponential distribution may give rise to few sites with 
a significantly larger values of $h_i$ resulting in a localized sample $(\bar{r}_s\approx0.39)$,
explaining the bimodal structure of the $P(\bar{r}_s)$ distribution for the speckle disorder
with $\sigma=0$. This mechanism {is further} reinforced by the presence of correlations 
in the speckle potential ($\sigma=1,2$). 

All in all, our results confirm that the inter-sample randomness plays a significant role in the 
ergodic-MBL transition both for the UR and speckle disorder. This suggests that both UR and speckle disorder
correspond to the same universality class of MBL transition, dominated by strong fluctuations 
{in sample-to-sample properties} and interplay of 
ergodic and localized grains. This is in sharp contrast to the MBL transition in quasiperiodic 
potential observable in current experimental setups where the system properties are much more uniform
\cite{Agrawal20}. 

\subsection{Participation entropies of eigenstates and multifractality}

While gap ratio statistics provides statistical information about 
eigenvalues of the model, additional information may be gained from 
eigenvector properties. Those may be {probed}
via e.g. participation entropies. Following \cite{Mace19b} we consider
 participation entropies  $S_{q}$ defined via the $q$-th moments
 of  wavefunction $\left| \Psi \right>$ following the exprespages = {036206},sion:
\begin{equation}
	S_{q} (\Psi)= \frac{1}{1-q}\ln\left( \sum_{i=1}^{N} \left| c_{i}\right|^{2q} \right),
	\label{eq:pe}
\end{equation}
where $c_{i}$ are the coefficients of wavefunction 
$\left| \Psi \right>$ in the basis state $\left| n \right>$, 
i.e. $\left| \Psi \right> = \sum_{i=1}^{N}c_{i}\left| n \right>$, $N$ 
is the dimension of Hilbert space. While providing supplementary 
information to that hidden in eigenvalues one should remember 
that the participation entropies are basis dependent (becoming trivial 
in e.g. the eigenbasis of the {Hamiltonian}). We shall consider the 
eigenbasis  of $S^z_i$ operators, equivalent to the basis of Fock states in the language of spinless fermions.
On the delocalized side
this basis is to a large extent unbiased. On the localized size,
since the so called local integrals of motion of the Heisenberg chain 
\cite{Serbyn13b,Huse14,Ros15,Imbrie16, Wahl17, Mierzejewski18, Thomson18} 
{can be thought of as dressed $S^z_i$ operators}
this basis is rather close to the eigenbasis
(for any reasonable measure of the basis distance \cite{Bengtssonbook}).
While this may be considered as a drawback, this choice assures that 
participation entropies are sensitive to the localization transition.
We considered only the lowest moments $q=1$ and $q=2${:
$S_{1} (\Psi)= -\sum_{i=1}^{N}|c_{i}|^{2} \ln|c_{i}|^{2}$ and 
$S_{2}(\Psi) = -\ln\left( \sum_{i=1}^{N} \left| c_{i}\right|^{4} \right)$ that are equal to 
to the Shannon entropy of $|c_i|^2$ distribution and logarithm of the inverse participation 
ratio (IPR), respectively. }
 
To probe the distributions of participation entropies 
we again consider $L=16$ where we have accumulated data for $4500$ disorder realizations 
for all cases studied. We take $300$ eigenfunctions corresponding to the middle of the 
spectrum around $\varepsilon=0.5$.
We show here the results for $S_2$ - the logarithm of IPR - the celebrated measure of localization studies in 
single particle physics \cite{Fyodorov95,Evers00} as reviewed in \cite{Evers08}.  
\begin{figure}[ht]
	\begin{minipage}[h]{0.8\linewidth}
		\centering
		{\includegraphics[width=1\linewidth]{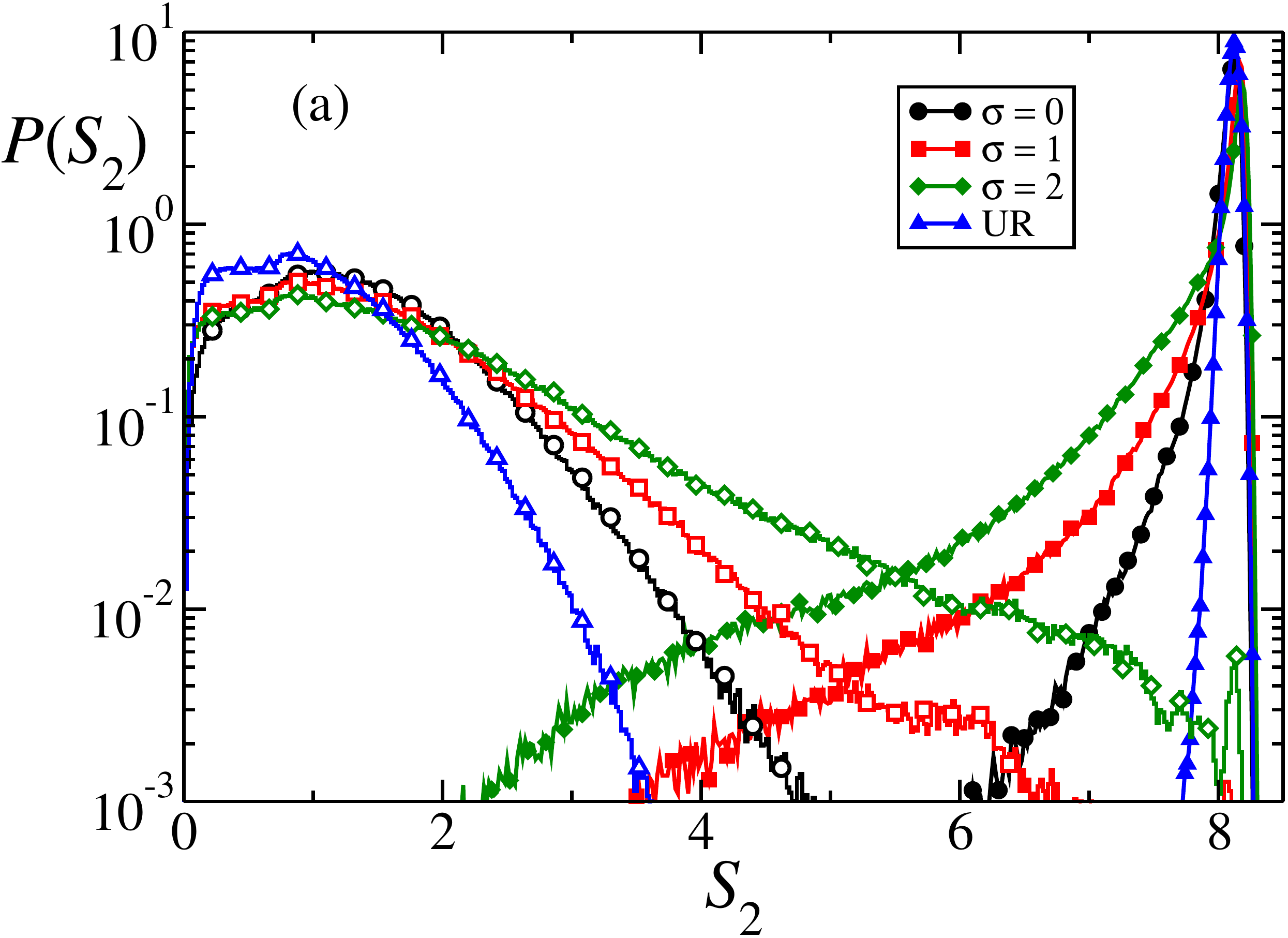} \\ (a)}
	\end{minipage}
	\vfill
	\begin{minipage}[h]{0.8\linewidth}
		\centering
		{\includegraphics[width=1\linewidth]{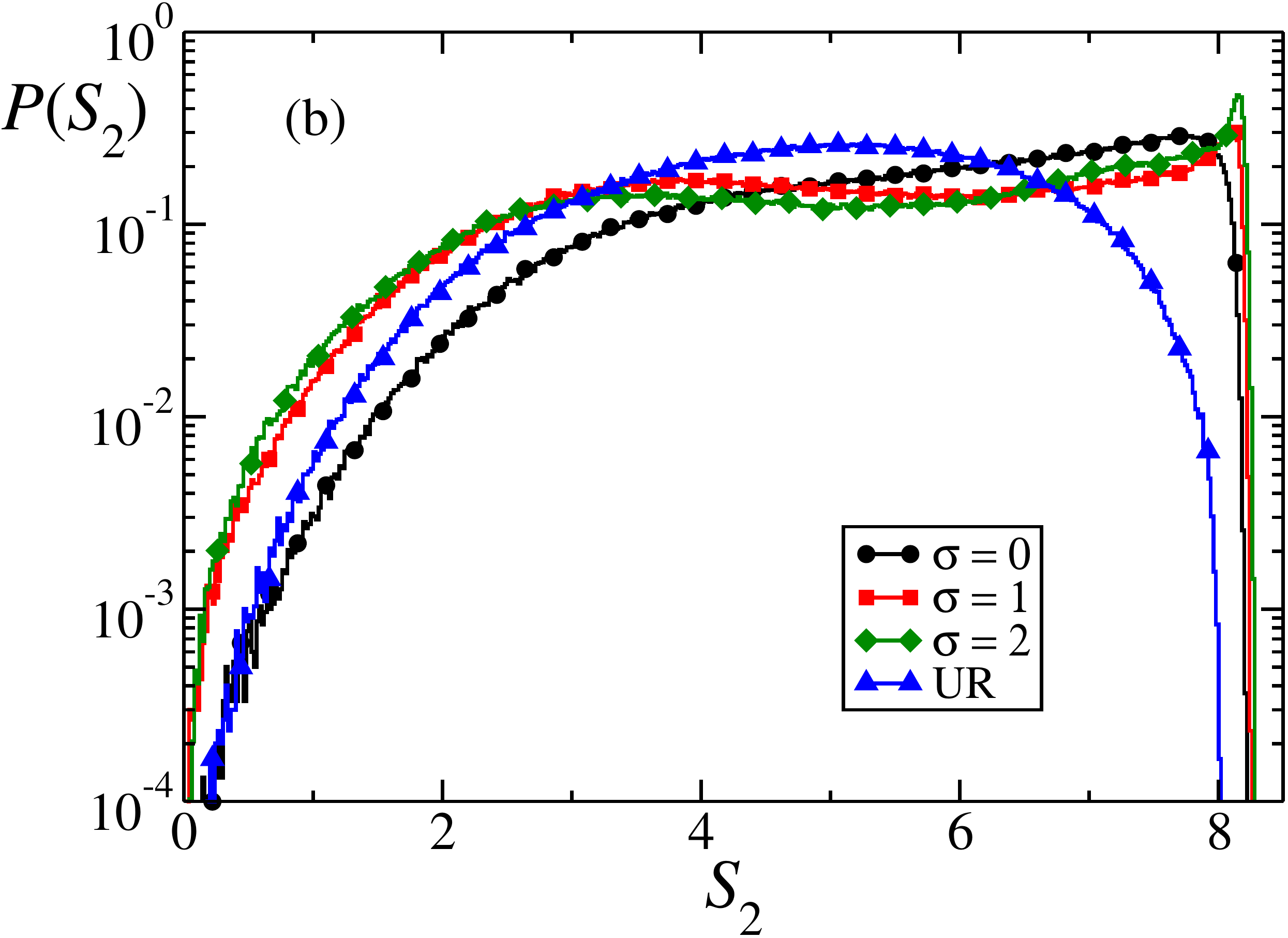} \\ (b)}
	\end{minipage}
	\caption{{The distribution of the participation entropy $S_2$ for deeply localized 
	(empty markers) and delocalized (filled markers) phases (a) and for the transition regime 
	(b). The data for delocalized phase are obtained for $w=0.15$, whereas the 
	ones for localized state are for $w=2.35$.  The transition regime corresponds 
	to maximal inter-sample randomness
	i.e. $w=0.5$ for the speckle and $w=0.6$ for the UR disorder.}}
	\label{fig:pe2}
\end{figure}

The $S_2$ distributions are shown in Fig.~\ref{fig:pe2}. Top panel 
compares {UR} disorder with the speckle for  localized and delocalized regimes
highlighting differences {in the two regimes}.
In the delocalized case, for {UR} disorder we observe
a narrow, almost symmetric gaussian-like distribution. This is not the case for speckle
potential despite the fact that $w=0.15$ lays deeply in the delocalized regime.
Distributions of $S_2$ show a pronounced asymmetry with a broad tail extending 
towards smaller values of $S_2$. The tail significantly grows with the speckle correlation
length. The presence of this tail indicates 
{relatively rare} situations where a partial localization occurs within the sample. A reversed 
trend is observed on the localized side. Here the participation entropy $S_2$ for an {UR}
disorder shows a characteristic
shape with a tail decaying as
$S_2^{-\alpha}$, for this value of $w$ {we find} $\alpha\sim2$. The uncorrelated speckle disorder
leads to a similar distribution but with,
again, the tail which decays more slowly. {The tail
gets heavier when the} {correlations in} speckle {disorder are introduced}.
 In the vicinity of the transition (see bottom panel of Fig.~\ref{fig:pe2}) 
the distributions become very broad for both types of disorder{, corroborating our claims about 
the same universality class for MBL transition in UR and speckle disorder, even in presence of a finite
range correlations in the latter case.} 
We also note that for the speckle 
potential an apparent excess of large $S_2$ corresponding to delocalized samples
 occurs.

The distribution shapes differ for UR and speckle disorder. 
A quantitative analysis is obtained by
finding the  $S_q$ scaling with the system 
size - here we follow closely the similar analysis performed
for RU disorder \cite{Mace19b}. It was shown that in that case to a good approximation 
$S_q=D_q\ln N +b_q$, where $N$ is the Hilbert space dimension of the system studied. 
{T}he eigenstates are multifractal if
{the fractal dimensions} $D_q$ differ among themselves. This is to be contrasted with $D_q=0/1$ for fully localized/delocalized case,
respectively. It was found that the Heisenberg chain in MBL regime 
possesses multifractal eigenstates. 

\begin{figure}[t]
	\includegraphics[width=1\linewidth]{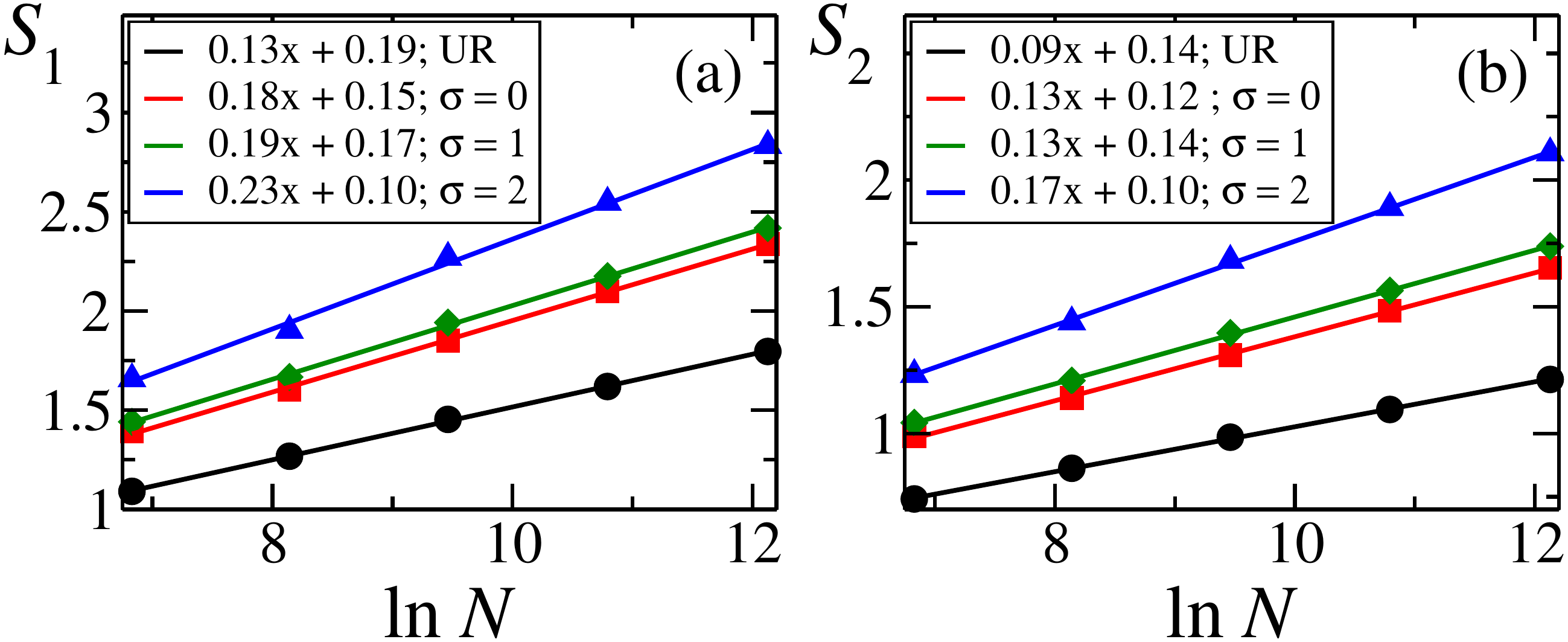}
	\caption{Finite size scaling of {participation entropies} 
	$S_1$ -- panel (a) and $S_2$ -- panel (b) for deeply localized phase ($w = 2.35$). 
	Data were acquired for 300 eigenlevels around $\varepsilon=0.5$ and 4500 disorder realizations.
	{The} fractal dimensions $D_q$ {(visible as coefficients 
	in front of $x\equiv \ln N$)} indicate multifractal character of eigenstates even in this 
	deeply localized regime {for all considered types of disorder}.  }
	\label{fig:sqscal}
\end{figure}

We {show that} this is also the case for the speckle disorder -- compare Fig.~\ref{fig:sqscal} where the scaling 
with the system size is studied in the deeply localized regime $w=2.35$. The speckle potential leads to slightly higher
multifractal dimensions in comparison to {UR} case as it  is already visible for $\sigma=0$. The  increase of speckle 
correlation length shows a further increase of multifractal dimensions revealing that localization is somehow ``weaker'' 
for the speckle potential at same, linearly rescaled disorder. Interestingly, observe, however, that the subleading $b_q$
coefficient remains positive for all cases considered (suggesting a localization 
as noted in \cite{Mace19b}).

\section{Time dynamics}
\label{time}

\subsection{Time evolution of the imbalance} 

While eigenvalues and eigenvector properties provide us with an understanding of the difference
between random and speckle potential they are not directly accessible in experiments. 
Standard MBL experiments 
\cite{Schreiber15,Luschen17,Lukin19,Rispoli19} consider instead the dynamics 
inferring the information from time evolution of 
appropriately chosen initial states. The first and simple conceptually approach \cite{Schreiber15}
considers the evolution of the density-wave like state with every second site occupied
and every second empty (for spinful fermions). Analogously, for the Heisenberg spin chain one may consider a 
N\'{e}el state {$|\psi(0)\rangle$} with
 spins up/down on even/odd sites, respectively (or vice versa). In time evolution starting from 
 {the} state $|\psi(0)\rangle$,
 the spin correlation function defined as
 \begin{equation}
 \label{eq:imb}
 I(t) = D \sum_{i=1}^{L}  \langle \psi(t) | S^z_i|\psi(t) \rangle   \langle \psi (0)| S^z_i |\psi (0)\rangle,
\end{equation}
where $D$ is a normalization constant (so {that} $I(0)=1$) is followed. 
The spin correlation function $I(t)$ for the N\'{e}el state is mapped, via Jordan-Wigner
transformation, to the difference of populations of spinless fermions at even and odd sites at time 
$t$, hence we  refer to it as an imbalance. 
While in the delocalized regime the imbalance rapidly decays to zero
in accordance with eigenstate thermalization hypothesis \cite{Rigol08, Alessio16}, in fully localized case,
after an initial transient, it saturates to a certain value dependent on the disorder amplitude. 
Theoretical and experimental studies \cite{Luitz16,Luschen17} addressed the time dependence of imbalance
also in the transition regime observing typically its power-law decay. This effect has been 
used to estimate the critical disorder for  MBL transition for large system sizes
\cite{Sierant18,Doggen19,Chanda20t,Chanda20m}.

\begin{figure} 			
	\includegraphics[width=1.0\linewidth]{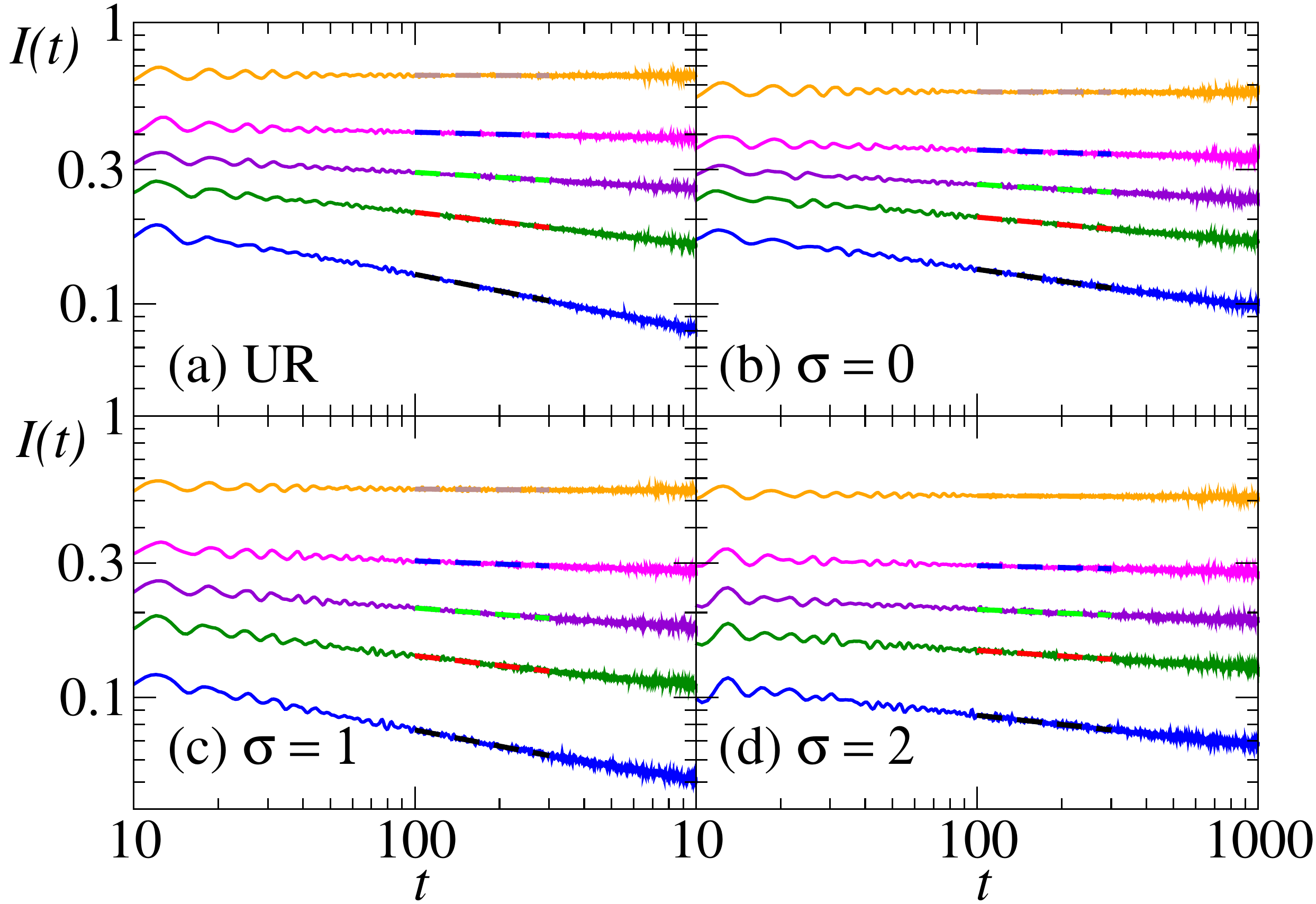}
	\caption{ The dynamics of imbalance for the system with random (a) and
	speckle potential (b)-(d) at disorder values (from bottom to top in
	each subpanel): $w=0.45$, $w=0.6$, $w=0.75$, $w=1$, $w=2$. The data are 
	obtained for N\'{e}el state by averaging over 600 disorder realizations
	for $L=20$ sites system. The $\alpha$ is extracted by power law fitting of 
	imbalance curves within the range $100 \leq t \leq 300$ (shown by dashed lines).
	}
	\label{fig:imbal} 
\end{figure}
The exemplary behavior of the imbalance \eqref{eq:imb} for the studied disorder 
types is depicted in Fig.~\ref{fig:imbal}. Results are obtained for $L=20$ system using
 Chebyshev propagation technique \cite{Tal-Ezer84,Fehske08}.  The curves indeed show the algebraic decay 
with $I(t)\propto t^{-\alpha}$ that persists to long times and disorder strengths. 
The algebraic decay describes well the time dependence of $I(t)$ for both UR and speckle disorder.

\begin{figure}[b]
 \includegraphics[width=1.0\linewidth]{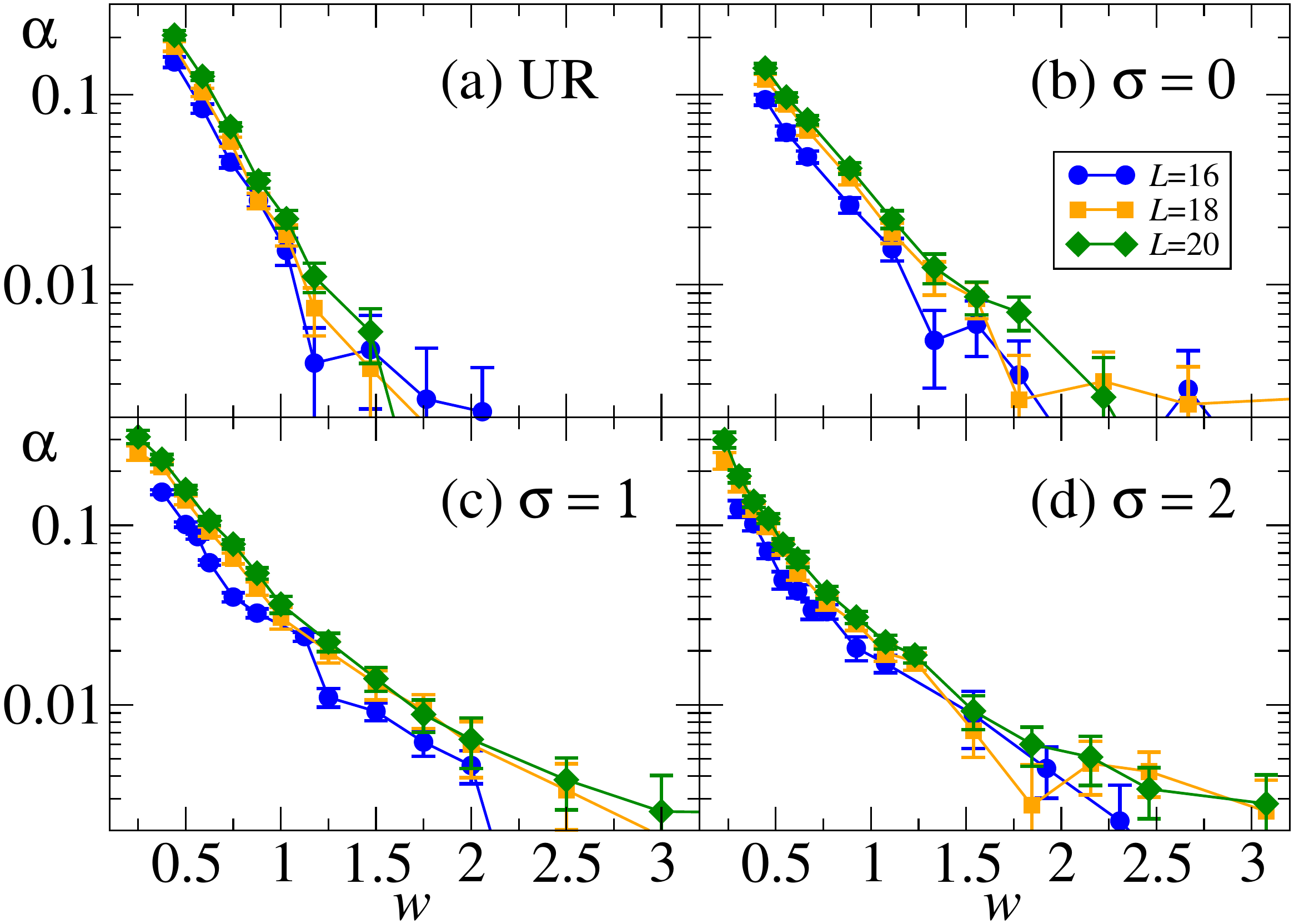}
 \caption{{ Exponent $\alpha$ of the spin imbalance decay as a function of disorder strength $w$,
 as extracted from fits for different system sizes for random potential (a), speckle potential with 
 $\sigma=0$ (b), $\sigma=1$ and $\sigma=2$. 
 The data are obtained for 500 realizations. Standard bootstrapping approach is used to estimate the errors shown in the figure.
 }}
  \label{fig:imbalpower} 
\end{figure}

Fig.~\ref{fig:imbalpower} shows the fitted exponents $\alpha$
as a function of the scaled disorder amplitude 
for  system sizes $L=16,18,20$. Firstly,
we observe, that the size dependence is {similar} 
for UR 
disorder {and} for the speckle potential, 
there is little difference between {exponents obtained for 
the available system sizes}. {A question that we leave for further studies 
is whether the slow increase in the exponent $\alpha$ observed for system sizes $ 20\leqslant L \leqslant 200$
for UR \cite{Doggen19,Chanda20t,Chanda20m} appears for the speckle disorder as well}. 
Another interesting observation is that the power $\alpha$ changes
differently with disorder amplitude
for various cases depicted in Fig.~\ref{fig:imbalpower}. The functional form 
of $\alpha(w)$ for UR disorder is well approximated by an exponential decay 
(a straight line in the lin-log plot). A similar dependence is apparent for an 
uncorrelated speckle potential, with one difference: the exponent $\alpha$ 
decreases significantly more slowly with $w$.  
The presence of correlations in the speckle disorder further enhances the slow decay of imbalance 
at large disorder strengths as we observe a non-zero exponent $\alpha$ even for $w>2$. 
This suggest that a 
linear rescaling of the disorder (by the critical disorder value) does
not fully compensate for
correlations in the disorder, the transition for larger correlation 
length  is ``broader'' with larger transition region. This parallels a 
similar observation made for participation entropies as well as the dependence
of the fractal dimensions on the speckle correlation length $\sigma$ {and can be linked to a formation
of small ergodic grains deep in the MBL phase as we show in the next section}.

\begin{figure}[b]
 \includegraphics[width=1.0\linewidth]{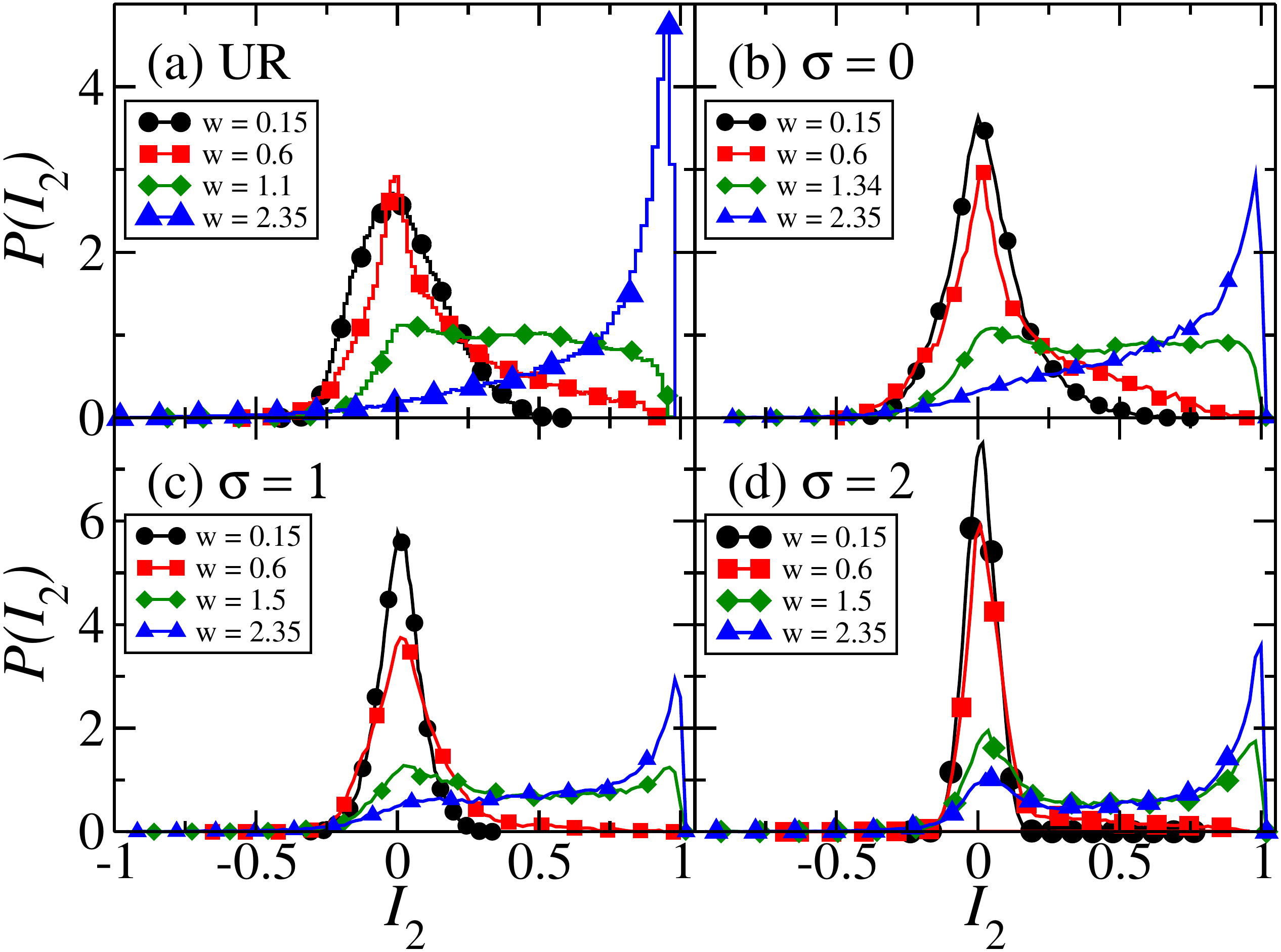}
 \caption{The distribution of local imbalance $I_2$ for various considered
 {types of disorder}.
 Observe the pronounced effect the speckle correlation has on that distribution.
 {The system size is $L=16$, the distributions are obtained for time $t=1000$ and from 
 more than $2000$ disorder realizations. }
  }
  \label{fig:locimb} 
\end{figure}

\subsection{Local imbalances}

Is it possible to reveal in a more pronounced way the different properties of MBL for UR and 
speckle disorders in time dynamics? After all those differences were quite strikingly visible in
 the {eigenvector} spectral properties w{as well as in} participation 
entropies. The answer is positive. 
What we need is a measure of {\it local} localization properties,
as {the} broad entropic distributions discussed above convincingly revealed the presence, even in the same sample, of
regions seemingly localized to a different degree. Such a measure may be constructed as a local imbalance
$I_2^k$, $k\in[1,L/2]$ involving spins on $2k-1,2k$ sites:
\begin{multline}
 \label{eq:imb2}
  I_2^k(t) = 2[\langle \psi(t) | S^z_{2k-1}|\psi(t) \rangle   \langle \psi (0)| S^z_{2k-1} |\psi (0)\rangle\\
 +\langle \psi(t) | S^z_{2k}|\psi(t) \rangle   \langle \psi (0)| S^z_{2k} |\psi (0)\rangle
 ].
\end{multline}
The global imbalance is just a sum of $I_2^k$. The
local imbalances provide information about local scrambling of spin
degrees of freedom.

We consider a set of local imbalances $\{ I_2^k(t_i)\}^{k=L/2-2}_{k=2}$ at times 
$t_i=1000-i$ (in the $J^{-1}$ units)
 {for $i=0,1,...,30$}.
Gathering the sets of local imbalances $\{ I_2^k(t_i)\}$ for 
a large number of disorder realizations we plot the resulting distributions of local imbalance,
$P(I_2)$, in Fig.~\ref{fig:locimb}. Consider first 
the distribution for 
UR disorder shown in panel (a): for large disorder the distribution is sharply peaked near its maximal unit value indicating
an almost  complete localization and {a good} memory of the initial state. On the contrary, for small disorder we 
observe a smooth gaussian-like profile centered at $I_2=0$ - a signature of a lost memory of the initial state. 
With increasing disorder this distribution sharpens, becomes asymmetric (as a total imbalance becomes positive) 
developing a tail at positive values of $I_2$. 
Around a critical disorder value the distribution of $I_2$
is  broad reaching the edge at unity. For speckle uncorrelated disorder -- Fig.~\ref{fig:locimb}(b) -- the curves
look similar although a careful inspection reveals that for the localized case the 
tail extending to small 
$I_2$ is higher than for UR disorder. The broadest distribution, extending almost uniformly between zero and unity
is obtained at 
slightly different $w$ value, in comparison to the UR case.
The picture changes for the correlated speckle
disorder. We observe a spectacular narrowing of distributions in the delocalized case. This may be easily
understood,
once some disorder value is chosen for a given site, the next correlated site has, with a large probability, 
a similar disorder value facilitating delocalization. Interestingly the broadest distributions (at values
of disorder shifting towards slightly larger values) develop 
local maxima at  $I_2=0,1$ showing the 
abundance of
completely delocalized as well as localized local imbalances. Correlated speckle disorder leads thus to 
the formation of localized and delocalized grains of small size. This behavior becomes even more pronounced 
at $\sigma=2$ when even for large $w$ corresponding globally to the deep MBL case, a 
noticeable maximum at $I_2=0$ still exists pointing
towards the existence of small regions that locally thermalize. 

Note the real resemblance between bimodal distributions observed for local imbalance with the similarly 
shaped characteristics of the sample averaged  gap ratio {shown in Fig.~\ref{fig:rs}}. The local imbalance 
allows us  to get a similar understanding of the system behavior as eigenvalue statistics not only
on the level of the global properties reflected by the imbalance but on a deeper, local level. Let us stress that the 
measurement of local imbalances requires a single site resolution - however in cold atomic systems as well
as in spin models such resolution is already achieved experimentally \cite{Gross17}.

\section{Conclusions}
\label{concl}
We investigate an optical
speckle field placed on top of a quasi one-dimensional optical lattice  
which allows us to go beyond the continuum approaches 
considered so far and to model the system within a tight binding description 
in which the speckle field gives rise to a on-site perturbation.
 The speckle disorder obtained in that manner has unique features enabling a control over 
its correlation length. On one hand, the speckle disorder allows us to study MBL transition
in an uncorrelated ``trully random'' disorder, as opposed to the routinely realized experimentally 
case of quasiperiodic potential which leads to MBL transition of different universality class \cite{Khemani17}.
On the other hand, the speckle field
 opens up the possibility of studying the influence of 
correlations in disorder on many-body localization transition. 
A specific exponential distribution
function of the uncorrelated speckle already leads to
certain differences in
the system behavior as compared to usually studied random uniform disorder, the 
effects become amplified when the speckle correlation length is increased. We 
observe more pronounced finite size effects that are, surprizingly, hidden in 
the reduced sensitivity of the system response to changes of its size. This 
suggests that  system sizes of few hundreds sites, available in current 
experiments in optical lattices may be needed to  reach the thermodynamic 
limit. MBL in the speckle potential has increased sample to sample variation 
as compared to random uniform disorder, as visualized in the gap ratio analysis 
as well as in the study of eigenvector properties. With increasing correlations the
critical regime of transition broadens.

Additional insights may be obtained from time dynamics. Global imbalance decay 
confirms the broadening of the transition while a local imbalance, a tool introduced in this work,
allows us to visualize the origin of the resistance against localization
observed in (particularly correlated) speckle potential. Apparently it 
favors creation, even for large disorder amplitude, of small grains that 
locally thermalize. That resembles the behavior expected of
rare Griffiths regions for uncorrelated disorder but, surprisingly, the 
correlations of finite range actually enhance their importance. Excitingly, such local 
imbalances seem to be readily accessible experimentally offering possible
experimental verification of the results presented.

 \begin{acknowledgments} 
We thank Dominique Delande for providing the code for the speckle potential and  discussions. The latter were also enjoyed with Titas Chanda. 
We acknowledge support by PL-Grid Infrastructure. This research has been supported by 
 National Science Centre (Poland) under projects  2015/19/B/ST2/01028 (P.S. and A.M.), 2018/28/T/ST2/00401 
 (doctoral scholarship -- P.S.) and 2019/35/B/ST2/00034 (J.Z.). {Partial support by the Foundation for Polish Science under Polish-French
Maria Skłodowska and Pierre Curie Polish-French Science Award is also acknowledged.}
{P.S. acknowledges support by the Foundation for Polish Science (FNP) through scholarship START.}

 \end{acknowledgments}

\appendix

\section{Appendix}

We present here additional numerical results for the model.
Those results, while not essential for the conclusions reached in the main text, 
supplement them with additional numerical evidence.

\subsection{Energy dependence of the transition and density of states}
\label{appDOS}

\begin{figure}
\includegraphics[width=0.45\linewidth]{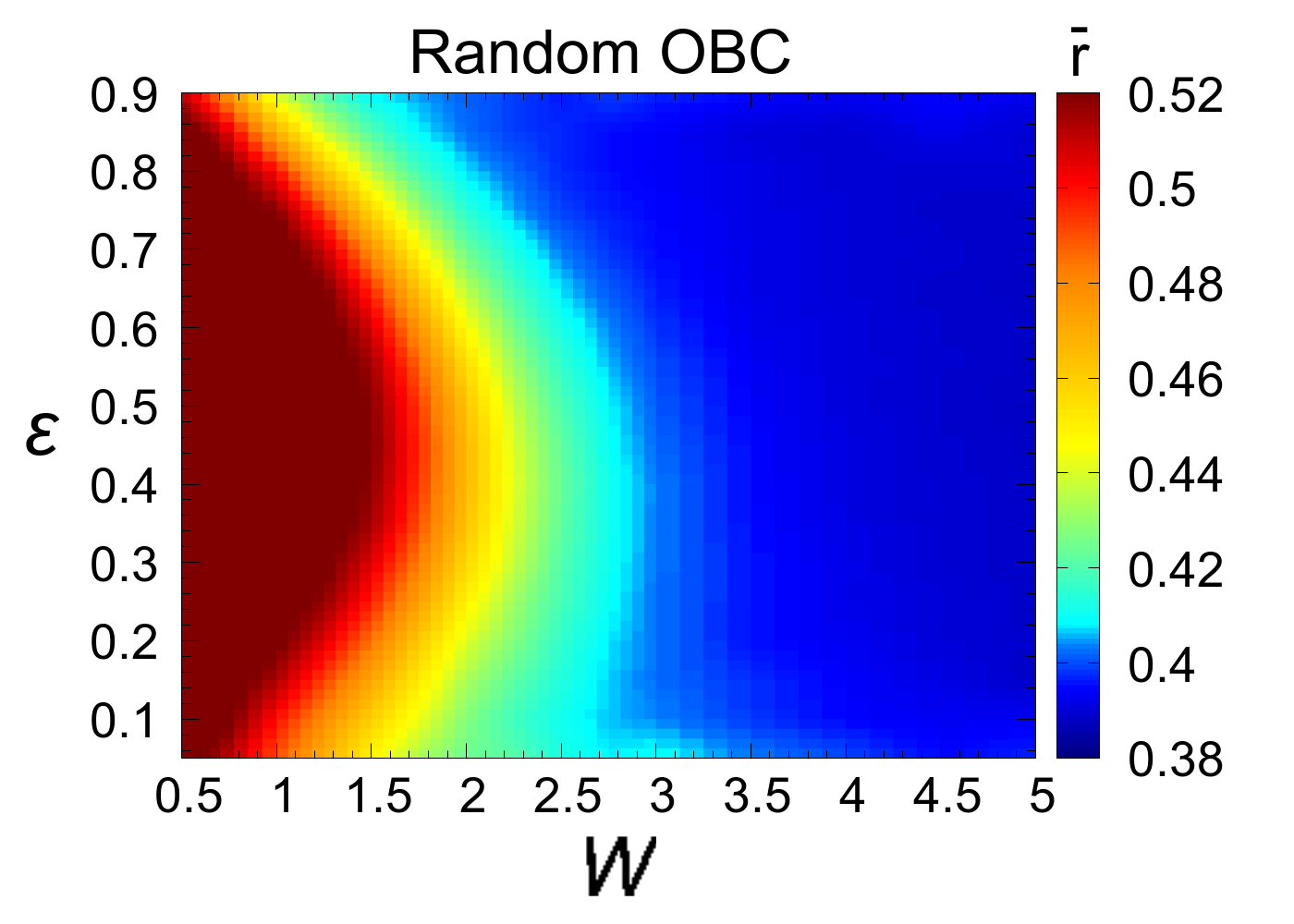}
 \includegraphics[width=0.45\linewidth]{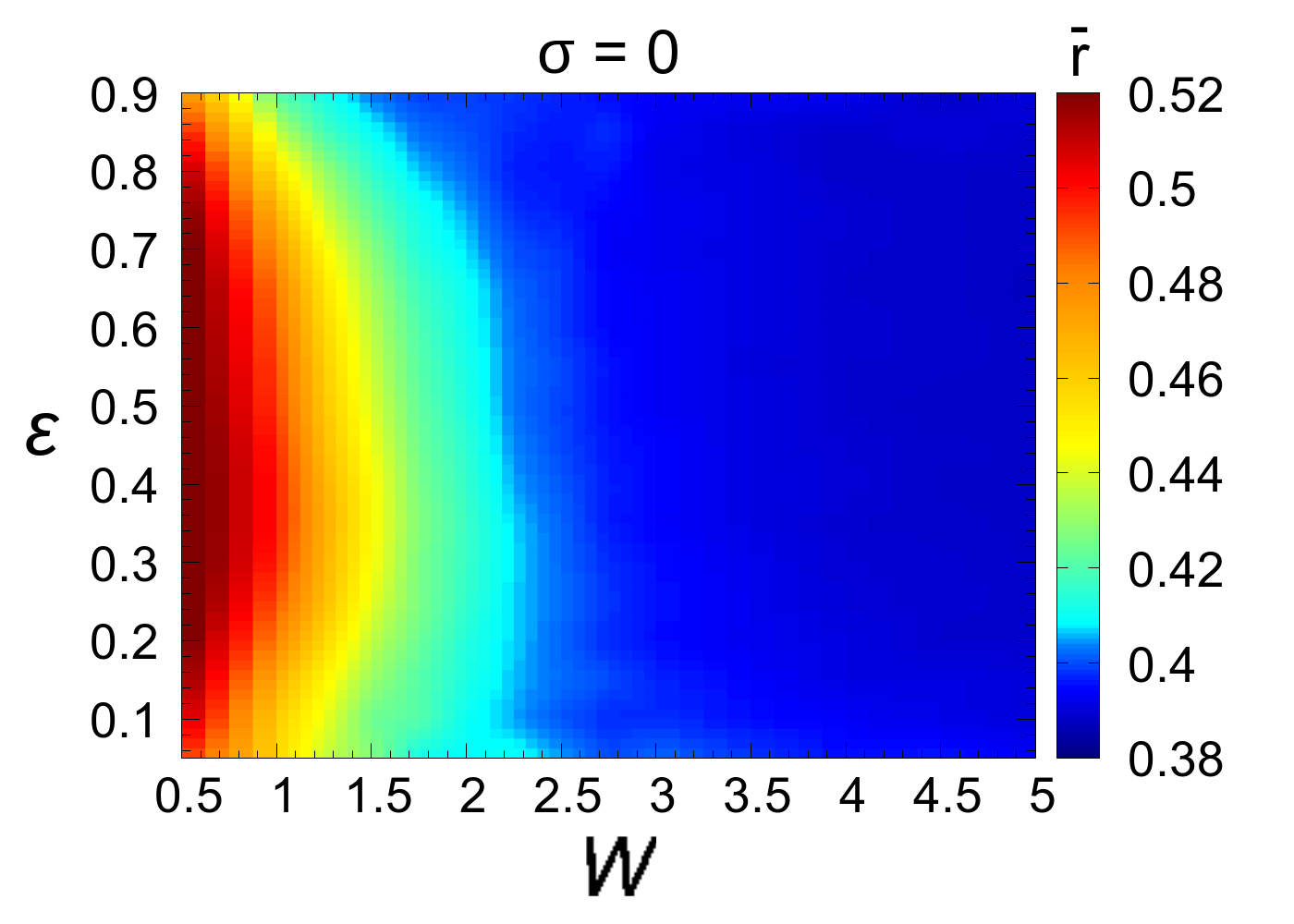}
 \includegraphics[width=0.45\linewidth]{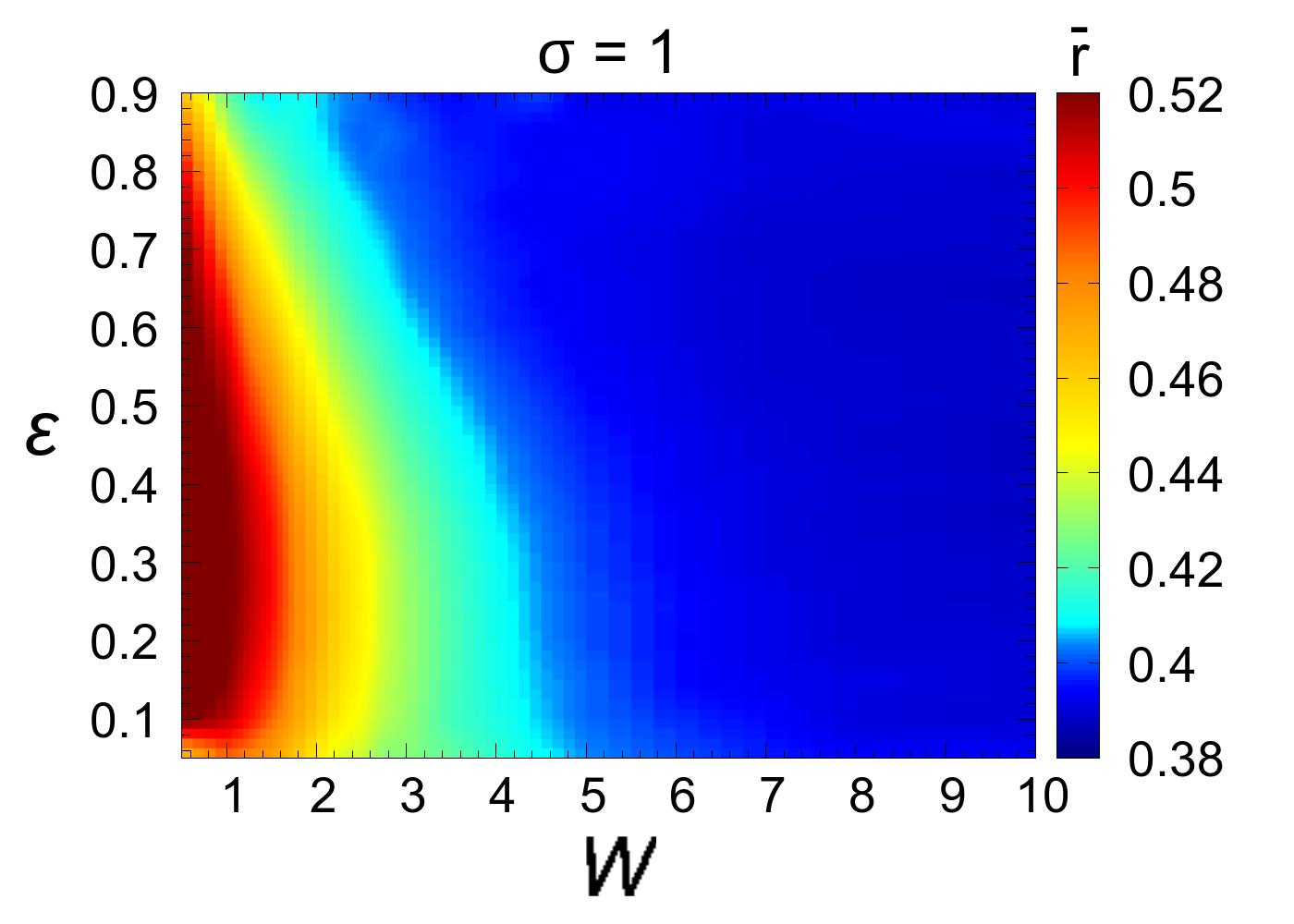}
 \includegraphics[width=0.45\linewidth]{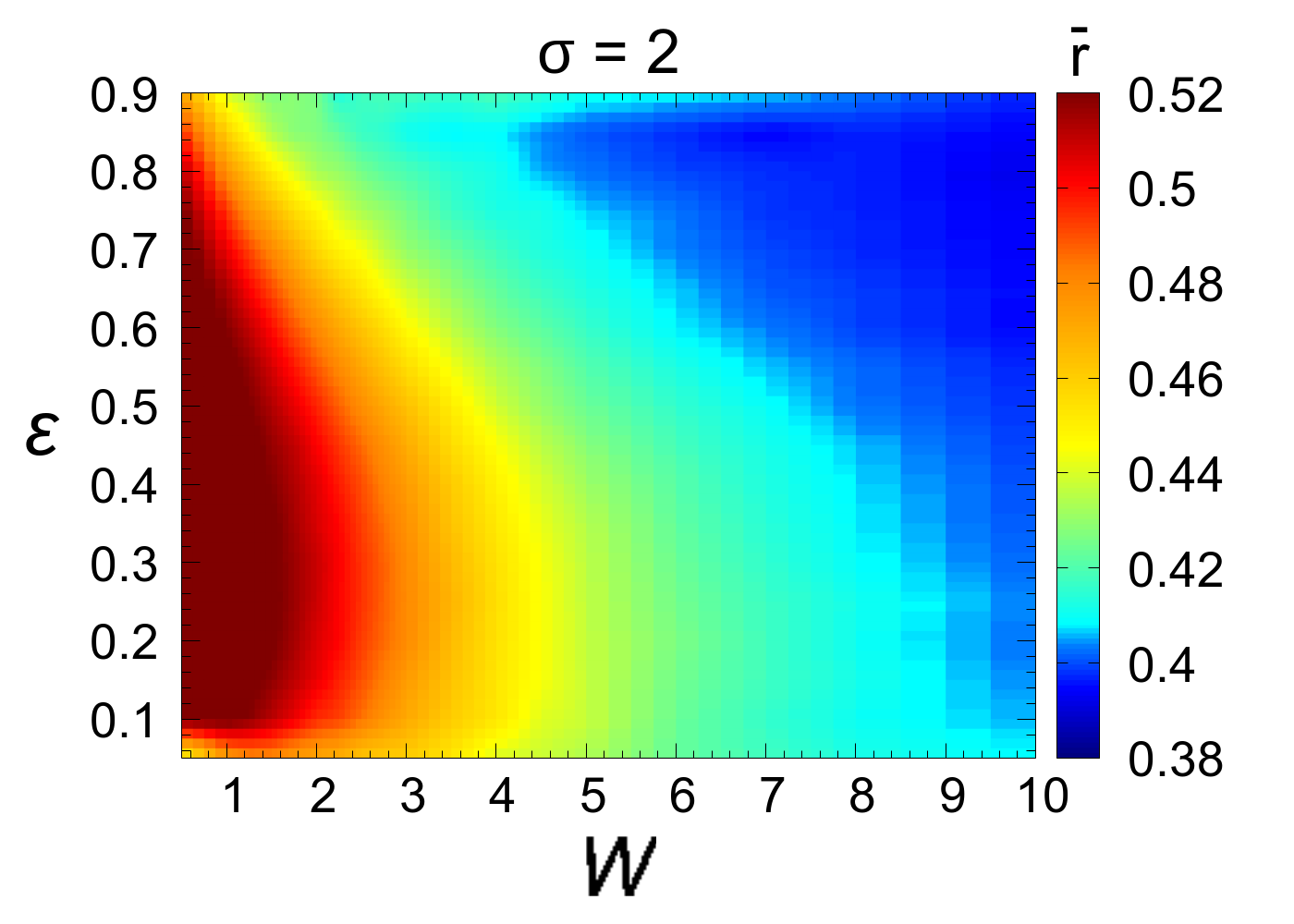}
  \caption{ {The mean gap ratio, $\overline r$, plotted as function of 
  disorder strength $W$ and the rescaled energy $\varepsilon$.} Observe that while uncorrelated, $\sigma=0$ 
  plot resembles, qualitatively, the situation for random uniform disorder 
  \cite{Luitz15}. For correlated disorder the lobe becomes asymmetric with 
  lower lying states being less localized.
  Data for $L=16$ Heisenberg chain.
  \label{fig:rbar} 
 }
\end{figure}

The {mean gap ratio } $\average r$  as a function of the disorder amplitude and the relative energy is presented in
Fig.~\ref{fig:rbar}. The latter is defined as $\varepsilon=(E-E_{min})/(E_{max}-E_{min})$
with $E_{min} (E_{max})$ being the lowest (highest) eigenvalue for a given disorder realization.
We take slices in $\varepsilon$ with size 0.05 obtaining 20 bins for energy. For
$\sigma=0$ i.e. an uncorrelated disorder we observe a rather symmetric in energy lobe 
resembling the one shown in 
\cite{Luitz15}. The correlation in disorder makes low lying states being more resistant 
to localization as clearly visible for $\sigma=2$ plot in  Fig.~\ref{fig:rbar}.

\begin{figure}[ht]
	{\includegraphics[width=0.49\linewidth]{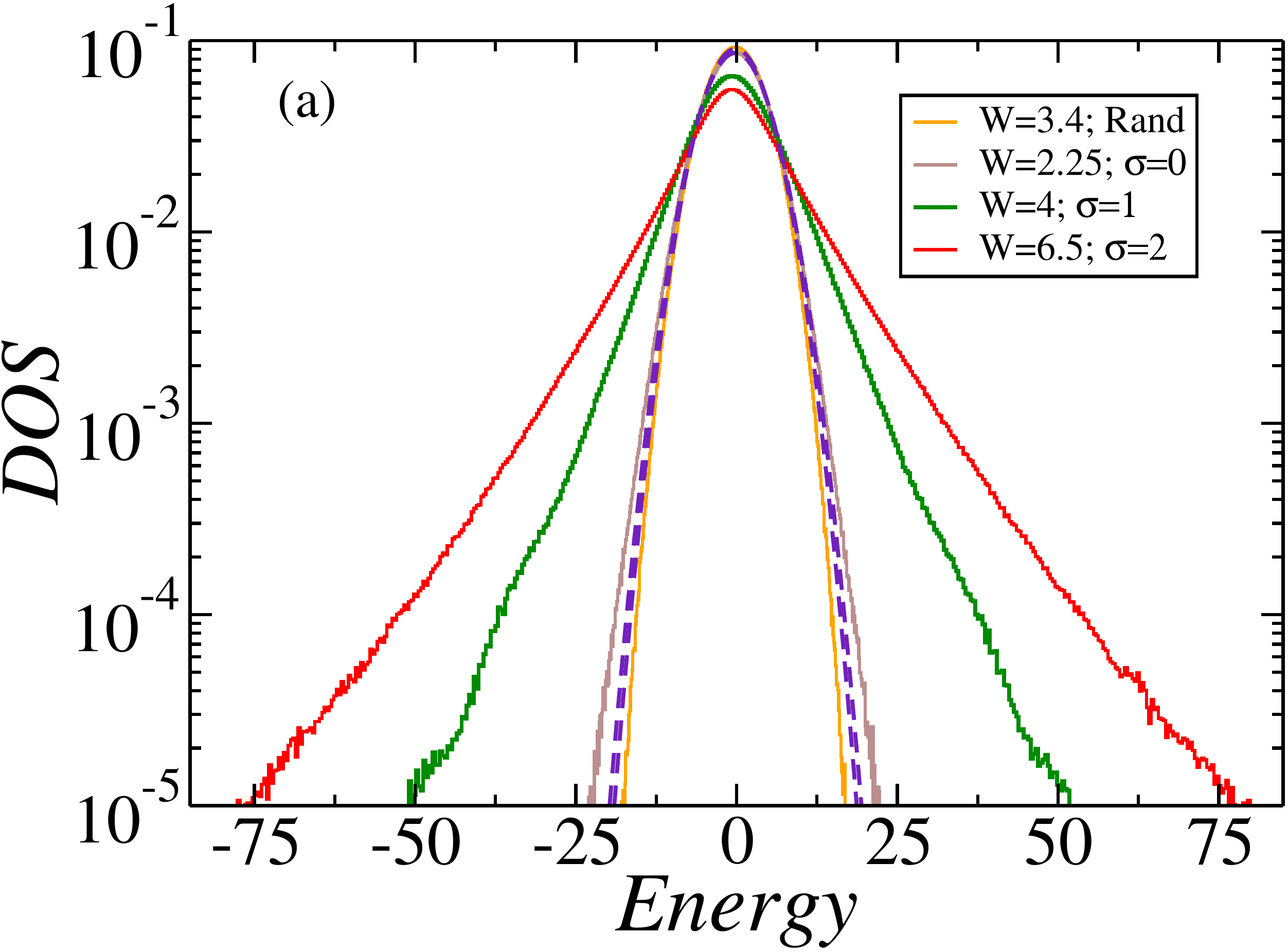}}
	{\includegraphics[width=0.49\linewidth]{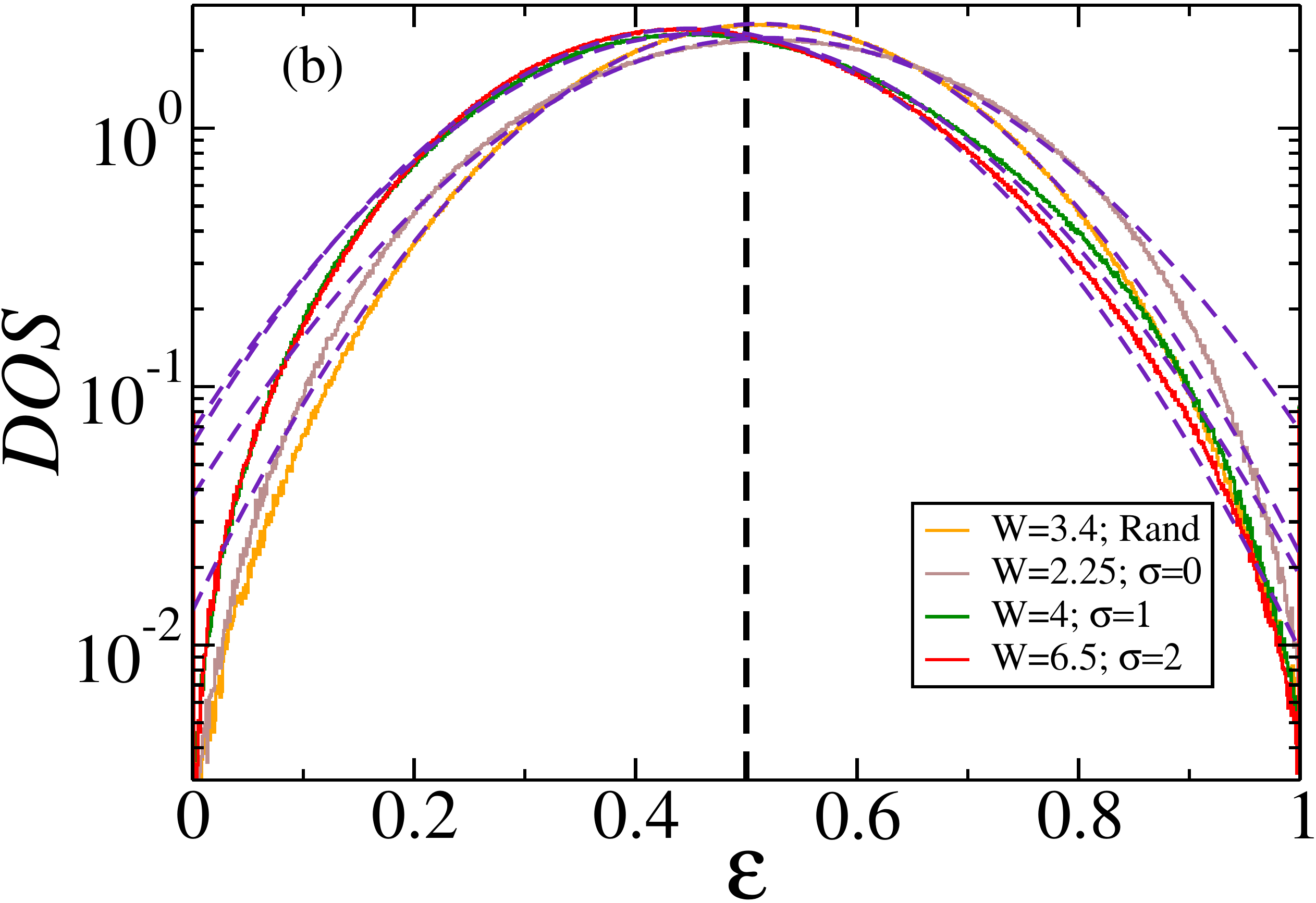}}
	\caption{The density of states [unscaled in (a) and scaled in (b)] for the system with the speckle potential. The dashed lines give Gaussian fits of the data. The data are obtained for $1000$ realization for the system of size $L=16$ and disorder values indicated in the figure - corresponding to our estimate of the critical values. 
	}
	\label{fig:dos1}
\end{figure}

This, at a first glance, surprizing behavior may be partially
explained by the energy dependence of the density of states - compare Fig.~\ref{fig:dos1}.
The raw (unscaled) density is perfectly symmetric with respect to origin. Scaling, performed
for each diagonalization separately, shifts slightly the maximum of the density of states 
to $\epsilon$ below $0.5$.
For a finite disorder value the maximum of the density concentrates around $\epsilon=0.4$ 
- that only partially explains the behaviour of the lobe for $\average r$ which, 
for a correlated disorder has a tip at around $\epsilon=0.25$.
Still, however, for $\sigma=2$ the maximum of DOS does not corresponds to the 
lobe in the energy gap ratio observed at $\varepsilon=0.25$. The central part 
of DOS distribution for scaled eigenenergies [Fig.~\ref{fig:dos1} (b)] is 
Gaussian for the system with the uniform random and the uncorrelated speckle 
potential (at least within the range $0.2\lessapprox \varepsilon \lessapprox 0.8$). 
For non-zero correlations in the speckle potential the range of DOS with approximately
Gaussian distribution  decreases with $\sigma$. The non-Gaussian character of DOS
distribution is more pronounced for unscaled eigenenergies where for correlated 
speckle potential the exponential tails are observed.

 \subsection{Correlations between participation entropy and the sample averaged gap ratio}
 
\begin{figure}[ht]
		{\includegraphics[width=0.85\linewidth]{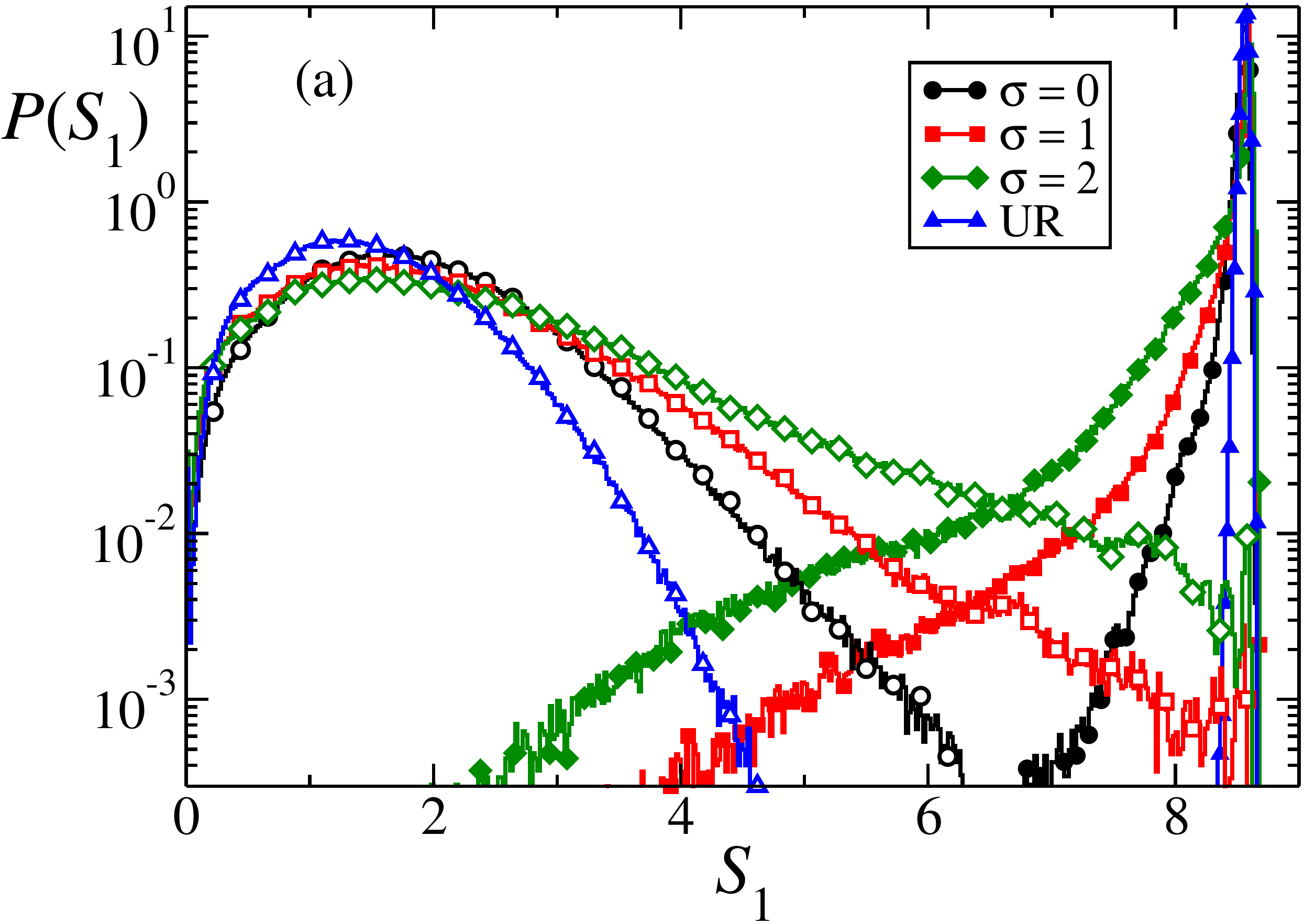} }
		{\includegraphics[width=0.85\linewidth]{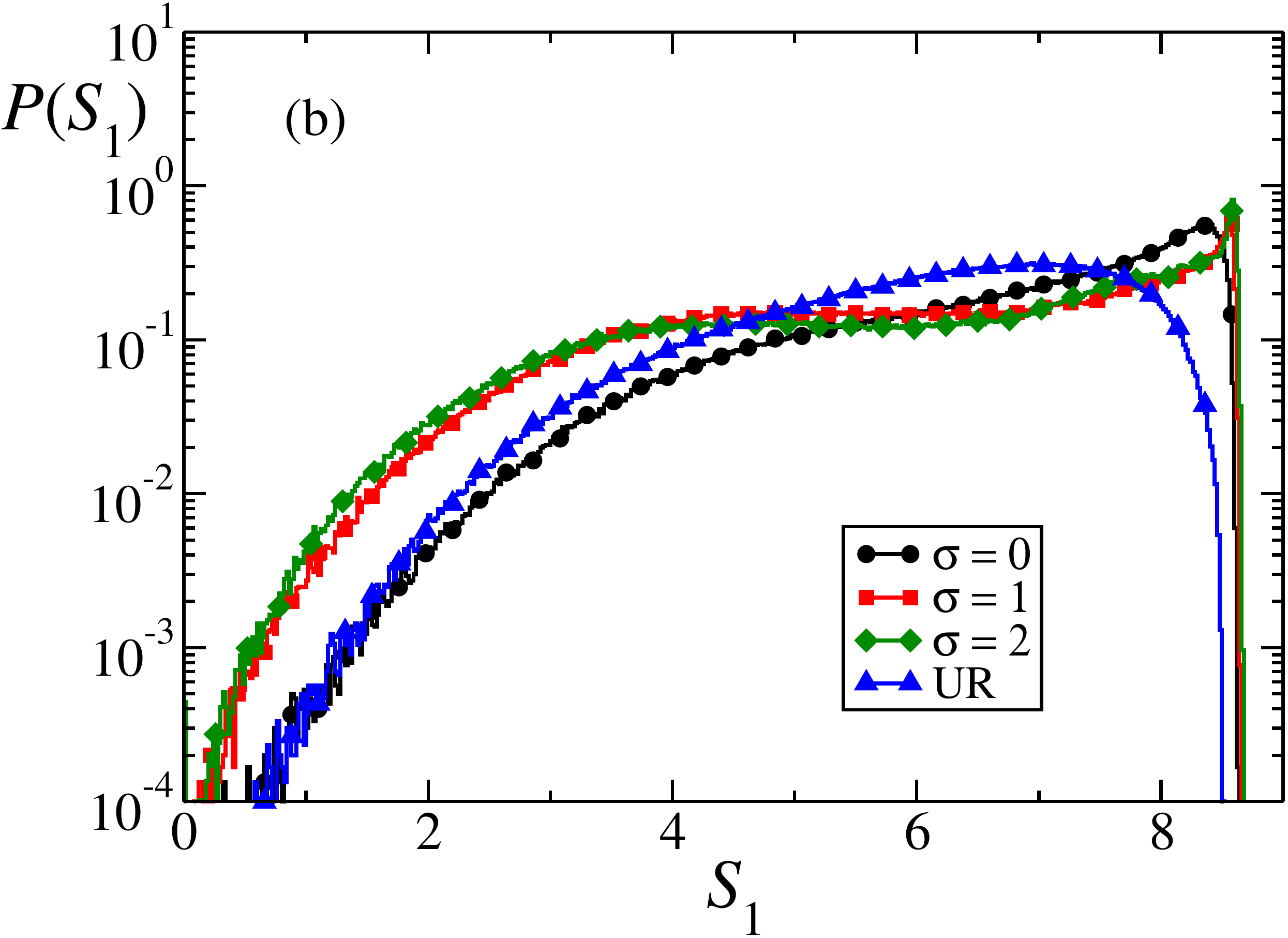} }
	\caption{{The Shannon entropy distribution for deeply localized (empty markers) 
	and delocalized (filled markers) phases (a) and for the transition regime (b). 
	The data for delocalized phase are obtained for $w=0.15$, whereas the ones for 
	localized state are for $w=2.35$. The transition regime corresponds to the 
	maximal $\bar r_s - \bar S_2$ correlations --compare Fig.~\ref{fig:sqrs} 
	i.e. $w=0.5$ for the speckle and $w=0.6$ for the random uniform potential.}}
	\label{fig:pe1}
\end{figure}

While in the main text we have shown the {participation entropy} $S_2$ distributions in different regimes, 
here we show, compare Fig.~\ref{fig:pe1} that similar picture we obtain for the Shannon 
entropy $S_1$. The relatively broad distributions
 obtained, in particular in the transition regime calls for comparing $S_q$ with gap 
 ratio distributions. 

  \begin{figure}[t]
	\includegraphics[width=0.9\linewidth]{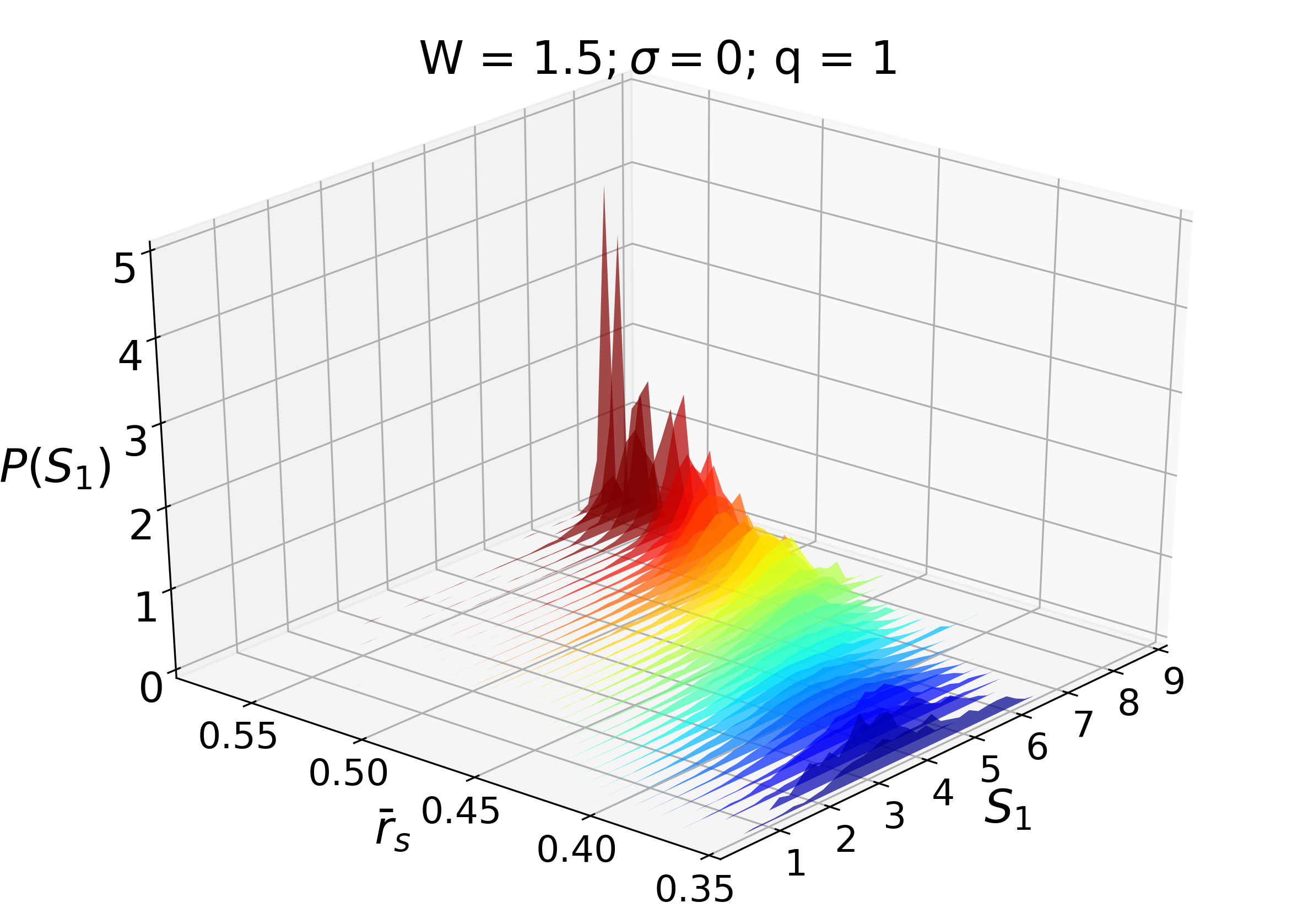}
	\caption{The partial distributions of the Shanon entropy $S_1$ as a function of $\bar r_s$ 
	for uncorrelated speckle potential ($\sigma=0$).The data are obtained for the system size $L=16$,
	300 levels around $\varepsilon=0.5$ and xx disorder realizations are taken into account. 
	The disorder $w=1.5/2.25=2/3$. Sharp distributions for ``delocalized'' samples shift and broaden 
	for lower $\bar r_s$ values. }
	\label{3D}
\end{figure}
\begin{figure}[t]
	\includegraphics[width=0.9\linewidth]{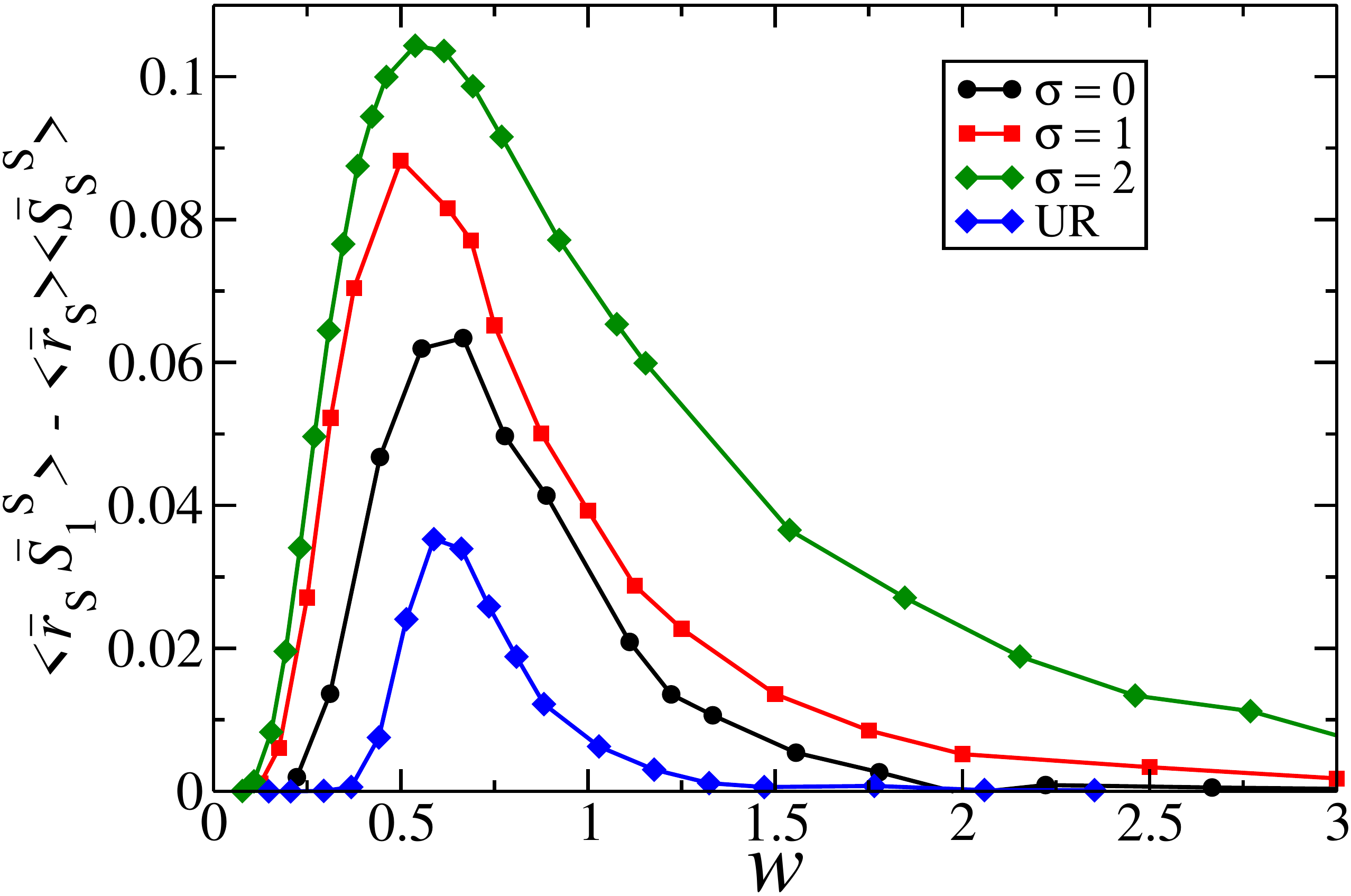}
	\caption{{The correlations between the sample mean gap
	ratio $\bar{r}_s$ and the mean Shannon  entropy $\bar S_1^s$ for 
	speckle and uniform random potentials (similar correlations are
	observed for $\bar S_2^s$). The data are obtained for 
	the system size $L=16$, 300 levels around $\varepsilon=0.5$ and 1000 disorder realizations.}}
	\label{fig:sqrs}
\end{figure}

 Having in mind the sample-to-sample randomness as 
 exhibited by $P(\bar r_s)$ one may consider the similar properties on 
 the entropy level. 
 { Fig.~\ref{3D} shows the distribution of $S_q$ obtained for disorder realizations
 with given sample averaged gap ratio $\bar r_s \pm \delta$, we take $\delta=0.006$.}
 {We observe a strong correlation between the center of the distribution of $S_q$ 
 and the value of the sample averaged gap ratio $\bar r_s$.}
 To quantify {those} correlations   we consider the 
 eigenvectors for 300 eigenvalues around $\varepsilon=0.5$, 
 find their {participation entropies} and average them for each disorder realization separately. 
 Let as denote such an average of for $s$- sample of $S_q$ values as 
 $\bar S^s_q$. The resulting correlation between $\bar r_s$ and 
  sample averaged Shannon entropy $\bar S^s_q$ is shown in Fig.~\ref{fig:sqrs}.
 The correlations are bigger for the speckle than for the random uniform 
 disorder and further increase with the correlation length of the speckle
 potential.  
 {The increase of correlations between the sample averaged quanities $\bar S^s_q$ 
 and $\bar r_s$ with correlation length $\sigma$ can be understood as 
 an effect of diminishing number of uncorrelated random fields $h_i$ in a given 
 disorder sample: for larger values of  $\sigma$, the potential fluctuations across a given sample are smaller 
 and hence properties of the sample, reflected either by $\bar S^s_q$ or $\bar r_s$, vary less yielding the larger  $\bar S^s_q$, $\bar r_s$ 
 correlation.  }
  For uniform disorder the maximum {of the $\bar S^s_q$, $\bar r_s$ correlation} occurs at the scaled disorder
 $w=W/W_c\approx 0.6$ shifting towards $w=0.5$ for the largest correlation
 length $\sigma=2$ considered by us. 
 With an increase of the system size the correlation maximum shifts  towards
 higher disorder values. One may speculate that, provided MBL persists in 
 the thermodynamic limit, 
 eventually the maximum shifts towards $w=1$ i.e. the critical point. The
 dependence of the position of the correlation maximum on system size was 
 checked for systems with $L=10$ to $L=20$ (not shown).  It was observed that
 the shift of the maximum is slower with increasing $\sigma$ supporting the 
 observation made for the gap ratio that finite size effects increase with 
 correlation length of the speckle potential.

\begin{figure}[ht]
	{\includegraphics[width=0.85\linewidth]{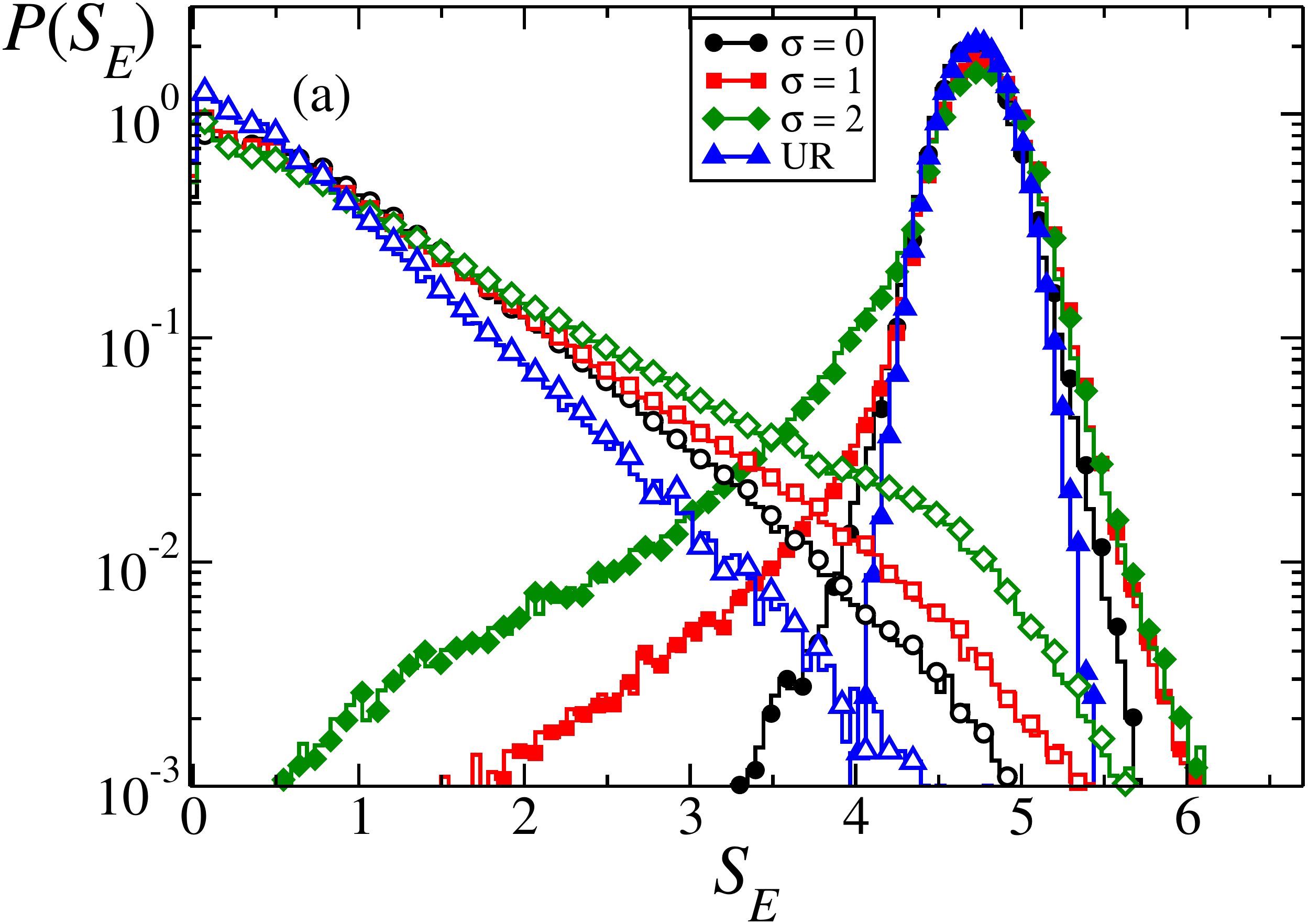}}
	{\includegraphics[width=0.85\linewidth]{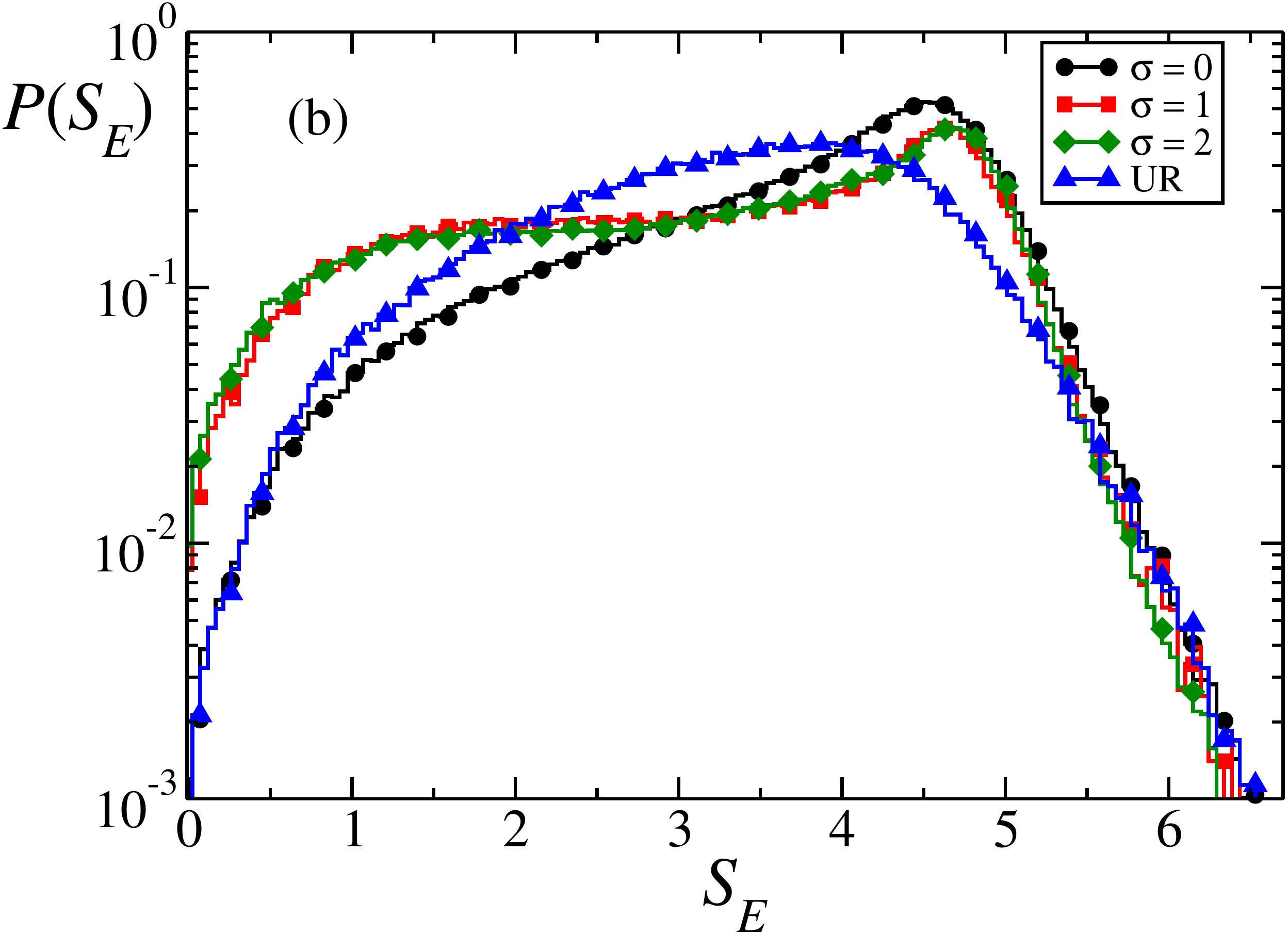}}
	\caption{The distribution of bipartite {entanglement entropy}
	for localized (open markers) and delocalized phase (filled markers) (a) and for 
	transient regime (b) for speckle (correlated and uncorrelated) and uniform random 
	potential. The disorder values are the same as in Fig.~\ref{fig:pe1}. The data are
	obtained for the system size $L=16$ divided in two equal parts ($L_A=8$), 300 levels
	around $\varepsilon=0.5$ and about 2000 realizations.}
	\label{fig:ee}
\end{figure} \subsection{Bipartite entanglement entropy of eigenstates}
 To complete the analysis of entropic properties let us consider a different measure, 
the bipartite entanglement entropy 
defined as $S_{e} = -\Tr\rho_{A}\ln\rho_{A}$ when the system is devided
into two parts $A$ and $B$. {The entanglement entropy}
is evaluated from Schmidt decomposition as follow
\begin{equation}
S_{e} = -\sum_{i=1}^{n}\lambda_{i}^{2}\ln\lambda_{i}^{2},
\label{eq:ee}
\end{equation}
where $\lambda_{i}$ are singular values obtained from decomposition.
The size of the subsystem A over which the partial trace $\rho_{A}$ is taken has beens
chosen as a half-size of the entire chain, i.e. $L_{A} = L/2$.

In an analogy to distributions {of participation entropy} 
we consider the distributions of entanglement entropy. 
The results are shown in Fig.~\ref{fig:ee} for delocalized and localized phase (Fig.~\ref{fig:ee}~(a)), 
and for transient regime (Fig.~\ref{fig:ee}~(b)).

The behavior of {entanglement entropy}
distributions is qualitatively similar to the ones for {participation entropy}. 
The UR potential leads to the Gaussian shape of{entanglement entropy} for delocalized phase whereas 
the  distributions for the system with speckle potential become asymmetric with
tails that {extends towards} small $S_E$ region, i.e. there are non-negligible number of values 
of $S_E$ that correspond to delocalized and localized phases. The overlapped area is
increased with increase of  correlation {length of the speckle
potential} $\sigma$. 

In localized phase the distribution of {entanglement entropy} seems exponential with the shape
independent {of} the nature of the disorder potential. In this case the slope 
(in the log scale) of the distribution is maximal for UR potential and
is decreasing for speckle potential especially with increase of the speckle correlation length.
In the transition regime the distributions are broad and become less
sensitive to the correlation length of the speckle potential.



%

\end{document}